\documentclass[12pt,letterpaper]{article}
\pdfoutput=1

\usepackage{amsfonts}
\usepackage{amssymb}
\usepackage{graphicx}
\usepackage{latexsym}
\usepackage{amsmath}
\newcommand\openone{\leavevmode\hbox{\small1\normalsize\kern-.33em1}}
\def\ee{\end{eqnarray}}
\renewcommand{\L}{\mathcal{L}}
\def\d{\partial}

\newcommand{\be}{\begin{eqnarray}}
\newcommand{\en}{\end{eqnarray}}
\newcommand{\bea}[1]{\left(\begin{array}{#1}}
\newcommand{\ena}{\end{array}\right)}

\newcommand{\tmop}[1]{\ensuremath{\operatorname{#1}}}
\begin{document}

%\begin{flushright} {\footnotesize IC/2007/001}  \end{flushright}
\vspace{5mm}
\vspace{0.5cm}
\begin{center}

\def\thefootnote{\fnsymbol{footnote}}

{\Large \bf Form Factor Dark Matter}
\\[0.5cm]
{\large Brian Feldstein,  A.~Liam Fitzpatrick, Emanuel Katz}

{\small \textit{
Physics Department \\
Boston University, Boston, MA 02215, USA}}

\end{center}

\vspace{.8cm}

\hrule \vspace{0.3cm}
{\small  \noindent \textbf{Abstract} \\[0.3cm]
\noindent

We present a dynamical alternative to inelastic dark matter as a way of 
reconciling 
the modulating signal seen at DAMA with null results at other direct 
detection experiments.
The essential ingredient is a new form factor which introduces momentum 
dependence in
the interaction of dark matter with nuclei.  The role of the form factor 
is to suppress events
at low momentum transfer.  We find that a form factor approach is most 
likely not viable 
in the context of the standard halo model, however it is consistent with 
halo models 
suggested by recent Via Lactea simulations.  As an example of possible 
form factors,
we present a class of models where the necessary momentum dependence arises 
from
interference of GeV mass gauge bosons coupling the dark matter to nuclei.  
At energies
relevant for direct detection experiments these models contain one or two 
additional parameters beyond the case of a standard WIMP.

\vspace{0.5cm}  \hrule
\def\thefootnote{\arabic{footnote}}}
\setcounter{footnote}{0}

\section{Introduction}
\label{sec:intro}

For several years now, the Dark Matter (DAMA) collaboration has observed 
an annual modulation in the nuclear recoil event rate at its 
NaI detector, and the significance of the detection is now at the $8\sigma$
level \cite{DAMALIBRA,DAMA}.  
It is striking that the modulation reaches its maximum within a week of 
June 2nd, when
the earth's motion into the galactic dark matter is at its greatest.
However,  if one supposes that the DAMA 
signal is to be explained by a
weakly interacting massive particle (WIMP) elastically scattering 
off of nuclei in the detector, then one predicts far
too many events to be seen at other experiments. The tension is particularly
acute at low nuclear recoil energies, where standard WIMP event rates are
expected to increase dramatically.

Although DAMA apparently cannot be reconciled with other experiments 
within the context of minimal WIMP models (see, for example, 
\cite{fairbairnschwetz}), 
this is not necessarily true in more general scenarios. Essentially, there is
no other experiment that is sensitive to the DAMA signal in a 
model-independent context. In particular, there are three crucial
aspects of the DAMA experiment which may be exploited in attempts at an
explanation. These are: 1) the masses of the particular nuclei used in the 
experiments, 2)  the fraction of the modulated vs. unmodulated 
event rate, and 3)
the particular ranges of nuclear recoil energies being probed.
Recently, inelastic Dark Matter (iDM) \cite{Neal1, SpencerNeal, JMR,
  lisapoland, Crucis, Winkler} has 
received a great deal of
attention for taking advantage of all three through
 a simple and elegant kinematic mechanism that easily arises in
specific particle physics models.  In fact, as the data has improved,
few if any other models are consistent with DAMA and all null experimental
results.  In this paper, we will present an additional class of models
as an alternative to iDM, which uses a dynamical rather than a
kinematic mechanism.

We will show how our mechanism can arise from a fairly straightforward model
where the dark matter interacts with the Standard Model by containing
constituents charged under new
dark forces that kinetically mix with the photon. In terms of
model-building, iDM is clearly simpler than our model. However, we wish
to emphasize that model-building dynamical dark matter form factors
can be an open direction for explaining DAMA, and there may be 
a variety of relatively simple alternative models. Thus, in the first part of this
paper, we will try to determine in a model independant way how well one can possibly do
with a form factor in terms of explaining all of the available data.  We will
see that the answer depends to an extent on the uncertainties
currently present in our understanding of the dark matter halo velocity
distribution.  The picture works somewhat poorly in the context of the
standard halo model, but much better with the results of the Via Lactea
numerical simulations.  In all cases we find a preferred dark matter mass
window, lying in a general range from about $30$ GeV up to about $50$ GeV, with
uncertainties coming from halo assumptions. 

The outline of this paper is as follows.  In section \ref{sec:review}, we
will review the general ingredients which go into predictions for direct
detection experiments.  Section \ref{sec:qoverlap} will discuss the
motivations for the form factor mechanism, as well as the model independent,
best-case-scenarios for the picture.  In section \ref{sec:models}, we will
present specific constructions which yield form factors with the 
appropriate momentum-dependence.
We will note that the models we present simplify considerably in the low energy theory
relevant for direct detection experiments.  In this limit, the relevant
parameters for the models become the mass of the dark matter particle,
the overall cross-section,
and the relative size of the $q^4$ term in the form factor
 compared to the $q^2$ term (as well as
possibly higher-power terms).  In this way,
the simplest theories we consider have exactly as many free parameters as
iDM.  Section \ref{sec:null} contains a summary of all of the present
experimental data relevant to our analyses, while in section
\ref{sec:results} we will use this data to constrain the paramaters of our
models, as well as give some examples of predicted spectra.  Conclusions and
future directions will be discussed in section \ref{sec:conclusions}.

\section{Review of Standard Elastic WIMP and Inelastic Dark Matter}
\label{sec:review}

Here we will briefly review the standard formalism for direct dark
matter detection experiments; for detailed reviews, see e.g. \cite{lewinsmith,
lewinsmithlong}.
The rate per unit detector mass per unit recoil energy for dark matter
nuclear scattering events is given by the general
expression
\be
\frac{d R}{d E_R} = N_T \frac{\rho_{\rm DM}}{m_{\rm DM}}\int_{v_{min}} d^3v f(\overrightarrow{v}) v \frac{d \sigma}{d
  E_R}.
\label{eq:drder}
\ee
Here $N_T$ is the number
of target nuclei per unit detector mass, and $\rho_{\chi}$ is the local halo
density of the dark matter particle, whose mass is $m_{\rm DM}$.
The rate depends on the local distribution $f(v)$ of the dark matter velocity
$v$ relative to the earth. The integral over the velocity
distribution depends on the nuclear mass and recoil energy only
through the minimum velocity $v_{\rm min} = \frac{1}{\mu}\sqrt{\frac{m_N E_R}{2}}$ 
that a dark matter
particle must have in order to result in a nuclear recoil energy of $E_R$,
where $m_N$ is the nuclear mass, and $\mu$ is the dark matter-nucleus 
reduced mass. 

The cross section for a standard WIMP on a nucleus with charge $Z$ and
atomic number $A$ is given by 
\be
\frac{d \sigma}{d E_R} = \frac{m_N}{2 v^2} \frac{\sigma_p}{\mu_n^2} \frac{(f_p
Z + f_n (A-Z))^2}{f_p^2} F_N^2(E_R).
\label{eq:xsec}
\ee
where $\mu_n$ is the dark matter-nucleon reduced mass, $\sigma_p$ is the dark
matter-proton cross section at zero momentum transfer, $f_n$ and $f_p$ are
the relative coupling strengths to protons and neutrons, and $F_N$ is the
nuclear form factor.  We will normalize our results to $f_n=0$ throughout,
corresponding to the specific models we present.  
A common parameterization of the nuclear form factor is the 
Helm form factor
\be
F_{\rm Helm}^2(q) &=& \left( 3 j_1 (q r_0) \over q r_0 \right)^2 e^{-s^2 q^2},
\ee
where $s=1$ fm, $r_0 = \sqrt{r^2 - 5 s^2}$, $r = 1.2 A^{1/3}$ fm, and
$q= \sqrt{2 m_N E_R}$ is the momentum transfer.  A more accurate
parameterization comes from the Woods-Saxon form
factor, which is the Fourier transform of the Two-Parameter
Fermi mass distribution $\rho(r) = {\rho_c \over e^{(r-c)/a}+1}$.
We will take the Woods-Saxon form factor for Iodine
($c= 5.593 \rm{fm}, a=0.523 \rm{fm}$), Germanium ($c=4.45 \rm{fm}, a=0.592 \rm{fm}$) and Tungsten 
($c=6.51 \rm{fm}, a=0.535 \rm{fm}$) \cite{Duda:2006uk, fricke}, and
use the Helm form factor for the remaining elements.  

Uncertainties in the form of the velocity distribution $f(v)$ remain
significant, as we will discuss in more detail, but a reasonable
starting point is that $f(v)$ in the galactic rest frame
is a Boltzmann distribution $\propto e^{-v^2/\bar{v}^2}$, with $\bar{v}
= 220 {\rm km}/{\rm s}$, 
and with a cut-off at the escape velocity $v_{\rm esc} = 550 {\rm km}/{\rm s} 
\pm 50 {\rm km}/{\rm s}$ \cite{vesc}. In our analysis, we will take
$v_{\rm esc} = $ 600km/s, since this end of the allowed range tends
to be more favorable for our models.  To obtain our observed
velocity distribution, the above expression must be boosted into the
Earth's rest frame.  The Earth is traveling through the
galaxy with velocity 
$v_e = v_s + $ $ 29.79 {\rm km}/{\rm s} 
( \cos(2\pi (t-t_{\rm JUNE})
\{ 0.262, 0.504, -0.823 \} + $ $ \sin(2\pi t- t_{\rm JUNE}) \{
-0.960, 0.051, -0.275 \} )$,
 where the sun's velocity is
$v_s = \{ 10, 225, 7 \} {\rm km}/{\rm s}$, and we use coordinates
with $\hat{x}$ pointing towards the center of the galaxy, $\hat{y}$ in the
direction of the disc rotation, and $\hat{z}$ towards the galactic north pole \cite{lewinsmith, lisapoland}.
The magnitude of the Earth's velocity through the galaxy 
is at a maximum on $t=t_{\rm JUNE}=$ June 2, causing a maximum
in the net flux of dark matter particles.  

The inelastic dark matter scenario involves a simple generalization of the above such
that a small amount of energy $\delta$, usually a mass splitting,
is lost in the scattering to the dark sector.
The only modification of the above is to increase the minimum
velocity by an amount $\delta/q$:
\be
v_{\rm min} &=& {q \over 2} \left( {1 \over m_{\rm DM}} +
{1 \over m_N} \right) + {\delta \over q}
\label{eq:vmin}
\ee
The additional contribution means that $v_{\rm min}$ gets pushed out to the tail
of the velocity distribution for smaller momenta $q$ and larger
dark matter masses $m_{\rm DM}$ than in the elastic case.  Its form
is also very significant: ${\delta \over q}$ grows at \emph{decreasing}
$q$, and in fact $v_{\rm min}$ has a minimum possible value 
$\sqrt{2 \delta \left( m_{\rm DM}^{-1} + m_{N}^{-1} \right)}$ that grows
with decreasing nuclear mass.  This is one of the main
original motivations for inelastic dark matter.

Event rates modulate the most over the course of the year
for recoils dominated by particles on the tail of the velocity
distribution.  DAMA is the only experiment looking at
a modulation signal, and ideally one way to try to reconcile DAMA with
null results at other experiments would be to choose the dark matter
mass sufficiently light(e.g. \cite{Feng:2008dz,Petriello:2008jj,Bottino:2008mf,
Foot:2008nw,Gondolo:2005hh}) in order to
make $v_{\rm min}$ at DAMA sufficiently large.
Unfortunately,
in the standard WIMP scenario, this means that the event rate
at DAMA should rise rapidly as the recoil energy decreases 
\cite{SpencerAaronNeal},
and this does not match the observed spectrum. Furthermore,
one faces immediate problems at the Germanium-based CDMS experiment,
where the lighter element and lower recoil energies 
imply lower $q$ and thus much
less kinematic suppression.

Fortunately, both of these problems stem from the sharp
rise in the predicted spectrum at low $q$, and may be addressed
by an additional form factor coming from the dark matter interactions,
as we now discuss.

\section{Overlap in $q^2$}
\label{sec:qoverlap}

\begin{figure}[t!]
\begin{center}
\includegraphics[width=\textwidth]{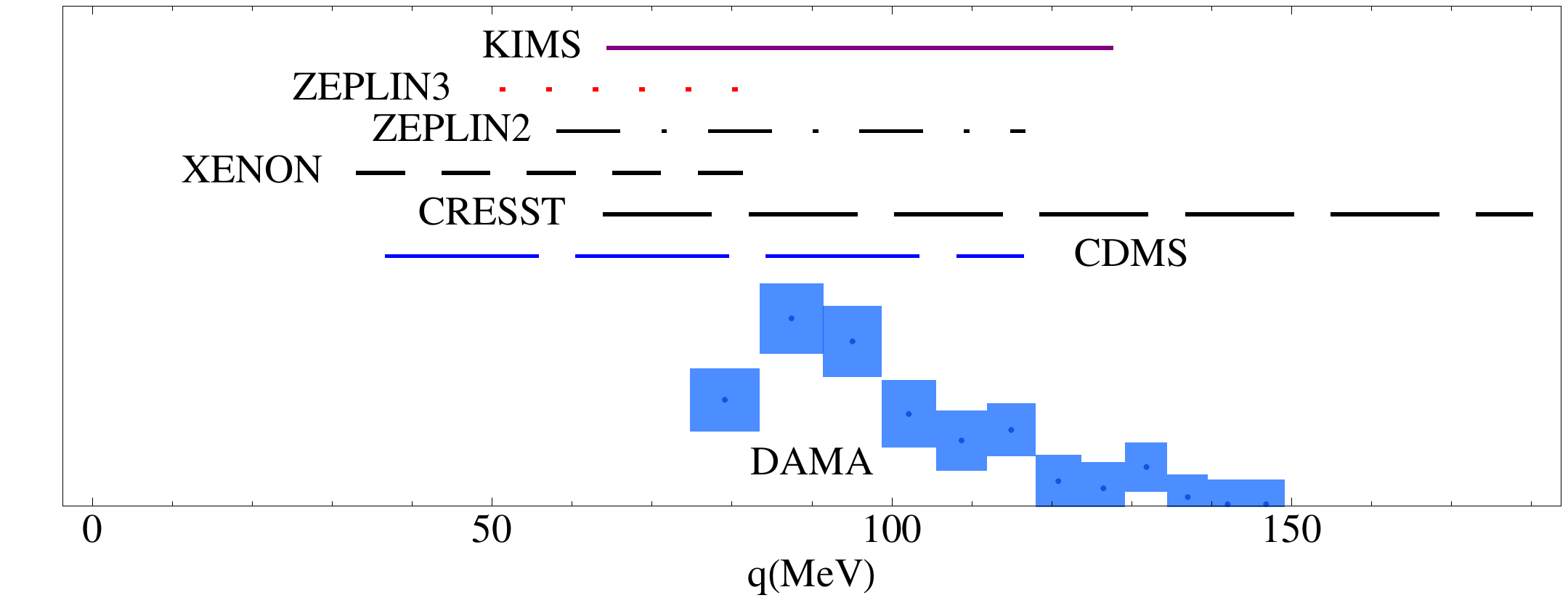}
\caption{ Overlap in $q$ of the DAMA signal with several
null experiments.  The height of the null experiments has no
particular meaning.    }
\label{fig:qoverlap}
\end{center}
\end{figure}

  The tension between the apparent signal at DAMA and the direct detection
experiments claiming null results becomes clearer through a comparison of
their overlap in momentum transfers $q$, as shown in
Fig. \ref{fig:qoverlap}.
In this paper, we will be considering scenarios in which the cross
section eq. (\ref{eq:xsec}) is multiplied by an additional function 
$F_{\rm DM}(q)^2$ coming from the dark
matter sector.  An important aspect of any such additional form factor 
$F_{\rm DM}(q)$
is that it does not depend on the nuclear mass and the recoil
energy separately, but only on their product $m_n E_R = q^2/2$.
As a consequence, any dynamical suppression at lighter nuclei can be
completely compensated for by looking at higher energies.  Perhaps more
importantly, a form factor cannot be used to suppress events at a
given experiment in an arbitrary way:  the DAMA modulating signal probes
values of $q$ between $q_i \sim 70$ MeV  and $q_f \sim 120$ MeV, and in this 
range one therefore
has significant constraints on the nature of the form factor.  Between $q_i$
and $q_f$, the form factor cannot be too small lest the DAMA signature
disappear, whereas outside of $q_i$ and $q_f$, there is no
model-independent lower bound.

  In practice, one would ideally like the
form factor to serve two distinct, but related purposes, 
alluded to above.  First, as
noted, one would like it to suppress the scattering rate at momentum
transfers outside of the DAMA range.
In addition to this, it should also serve to flatten the spectrum of events 
observed at DAMA
compared to the steep rise at low recoil energies 
predicted by a standard light elastically scattering WIMP.
Both of the above purposes may be served
simultaneously by a form factor which falls off appropriately at \textit{low} momentum
transfers.

Now, a key point is that the events seen by DAMA between $q_i$ and
$q_f$ essentially lead to a direct prediction (up to modulation fraction) for
the events to be seen at other experiments within that same range of momentum
transfers.
These predictions are more or less independent of the choice of form factor,
and it is therefore not immediately obvious whether they alone are enough 
to rule out
form factor dark matter as an explanation for DAMA.  The most basic question
we must answer is thus the following:  does there exist any function $F(q)$
for the form factor -
which we may take to be zero outside of the range $q_i < q < q_f$ - which 
allows for the DAMA modulating signal, but which does not overpredict the
number of events to be seen between
$q_i$ and $q_f$ at other experiments?

\begin{figure}[t!]
\begin{center}
\includegraphics[width=0.5\textwidth]{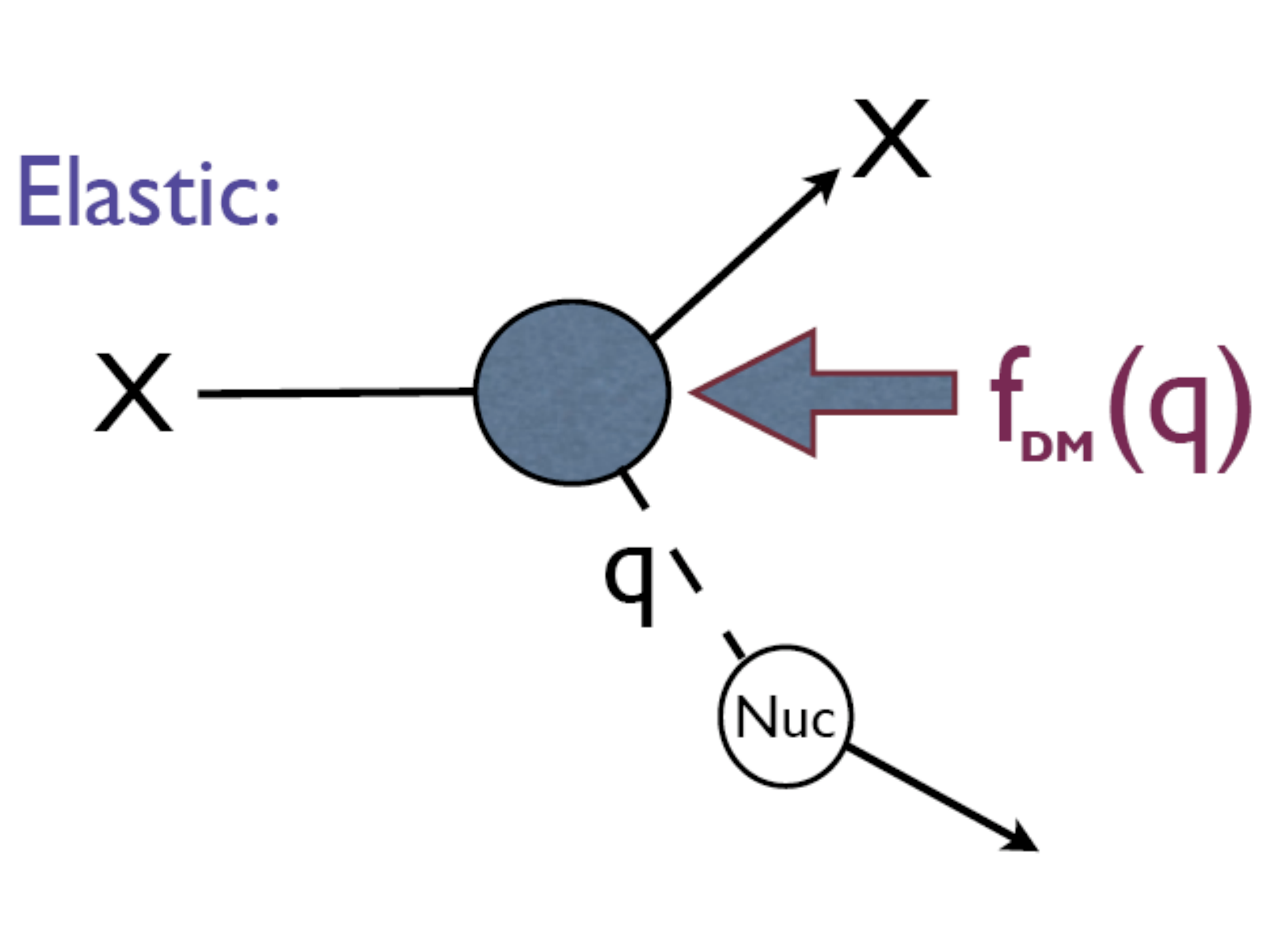}
\caption{ Schematic Feynman diagram for a form factor.    }
\label{fig:schematic}
\end{center}
\end{figure}

\begin{figure}[t!]
\begin{center}
\includegraphics[width=0.5\textwidth]{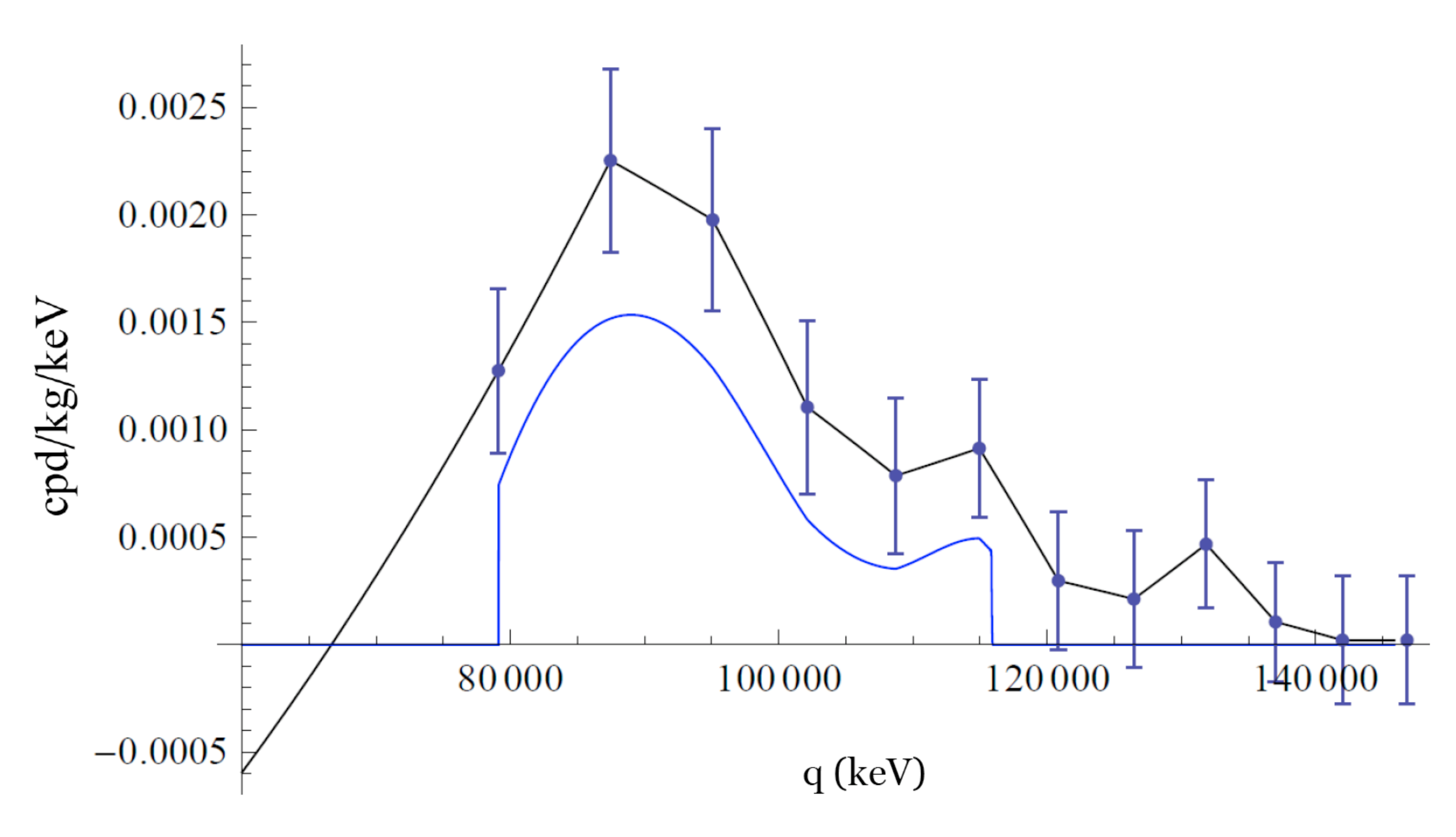}
\caption{ An example of a spectrum from an idealized form factor. The
discontinuous drop at low energies is by construction, to suppress
all events outside of the DAMA signal range, while the discontinuous
drop at high energies arises from the fact the escape velocity
$v_{\rm esc}$ cuts off the phase space beyond that point.}
\label{fig:bestformfactorexample}
\end{center}
\end{figure}

Later we will consider explicit models that give rise to form factors,
but for the moment we would like to answer this question while being as 
agnostic as possible about the model-building aspects.  
Thus we will begin by working with a physically unmotivated form factor,
chosen solely with the goal of fitting the DAMA observed spectrum while simultaneously 
being consistent with the null experiments.  To achieve this,
we will construct a form factor to explicitly 
put the signal just below the
1$\sigma$ error bar at DAMA, bin-by-bin\footnote{More precisely,
we construct the signal to be 80\% of the signal-minus-$1 \sigma$
rate.}.    Furthermore, outside the range of the DAMA signal 
(i.e. below $q=80 $ MeV), we set the form factor to zero.  
An example is shown in Fig. \ref{fig:bestformfactorexample}.
To evaluate the consistency of this form factor with experiment,
we calculate the probability of the low number of potential signal events
at CDMS and CRESST-II using the $pmax$ method \cite{SpencerNeal, Yellin}, 
which is based on the number of events predicted to have been seen in the gaps between 
potential events \footnote{
We present in detail our assumptions about the data from CDMS and
CRESST-II in section \ref{sec:null}.}.
We also calculate the $\chi^2$ of the fit at
DAMA, which changes with dark matter mass and halo parameters
due to the finite escape velocity.  In particular, sufficiently large
recoil energies necessarily have velocities above $v_{\rm esc}$,
and thus the signal is killed by phase space regardless of
the size of the form factor.  This tends to rule out small dark
matter masses.  We will take a fairly lax criterion for the
probability of the DAMA fit in the lowest 12 bins, calculating
only the goodness-of-fit for $\chi^2$ with 10 degrees of freedom.
Of course, currently the form factor has a very large number of
free parameters, but we are imagining the best-case
scenario where one has constructed a model that automatically
gives rise to the desired shape, and the only free parameters
are the dark matter mass $m_{\rm DM}$ 
and the overall size of the cross-section $\sigma_n$.

The range of masses available depends highly on the choice of dark matter halo
model, so let us now take a moment to describe our approach to
this additional uncertainty.
As we have mentioned, a commonly used simple model is
a Maxwell-Boltzmann distribution for the dark matter velocities
with a cut-off $v_{\rm esc}$ in the galactic rest frame.  
We will not attempt a systematic analysis of the effect of 
halo uncertainties, though in light of their effect we believe 
such an analysis would be very useful. Instead, we will
follow \cite{JMR} and consider two additional halo models,
VL$_{220}$ and VL$_{270}$,
based on the Via Lactea-I simulation \cite{vialacteaI}. 
 Such an analysis is conservative
in that halo parameters are notoriously difficult to constrain
and a thorough analysis marginalizing over all halo parameters would
open up more of the dark matter parameter space\footnote{We emphasize
that the main effect of the Via Lactea halo models will be due to their
tighter velocity distributions, and similar qualitative effects would
result from Maxwellian distributions with smaller RMS velocities.}.

The VL$_{x}$ models are based on a parameterization of
the radial and tangential velocity distributions \cite{fairbairnschwetz,JMR}
in the galactic rest frame, 
fit to Via Lactea simulation results:
\be
f(v_R) &=& {1 \over N_R} \exp \left[ -\left( v_R^2 \over \bar{v}_R^2 \right)
^{\alpha_R} \right] \\
f(v_T) &=& {2 \pi v_T \over N_R} \exp \left[ -\left( v_T^2 \over 
\bar{v}_T^2 \right)
^{\alpha_T} \right]
\ee
with $\alpha_R = 1.09, \alpha_T=0.73, \bar{v}_R= 0.72 \sqrt{-U(r_0)},
\bar{v}_T = 0.47 \sqrt{-U(r_0)}$, where $U(r_0)$ is a normalization
parameter.  The resulting $\sqrt{-U(r_0)}$
around $r_0 \approx $ 8 kpc in the Via Lactea-I simulation
 was 270 km$/$s; taking this value
gives the VL$_{270}$ model.  If one instead sets $\sqrt{-U(r_0)}$ to
the value 220 km$/$s in order to obtain a somewhat tighter distribution, then
the result is
the VL$_{220}$ model. Given the uncertainty in the
halo parameters and the difficulty in simulating the Milky Way
(e.g. including baryons), we believe
VL$_{220}$ is within the allowed halo model
parameter space and is important to include.

\begin{figure}[t!]
\begin{center}
\includegraphics[width=0.48\textwidth]{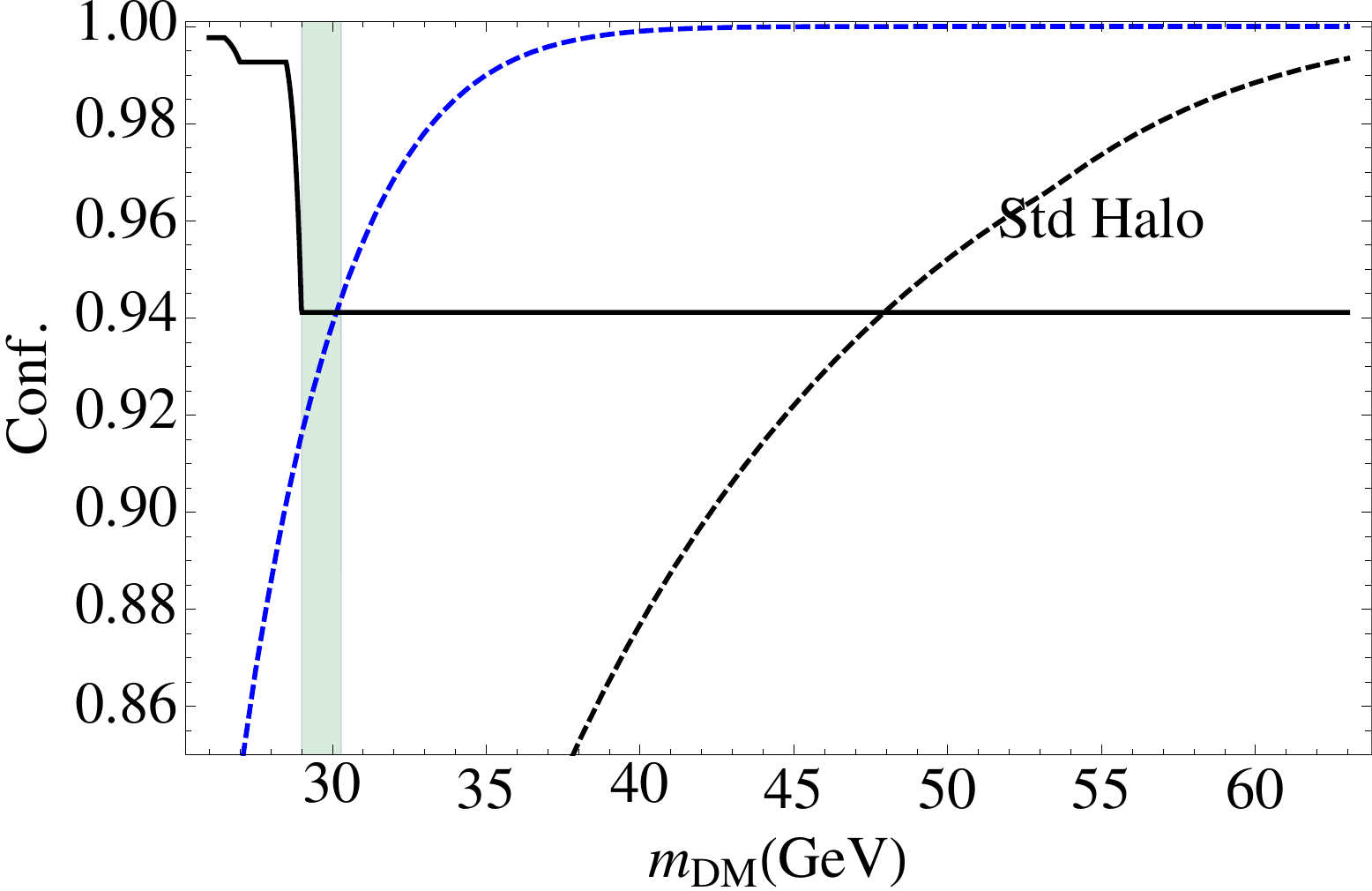}
\includegraphics[width=0.48\textwidth]{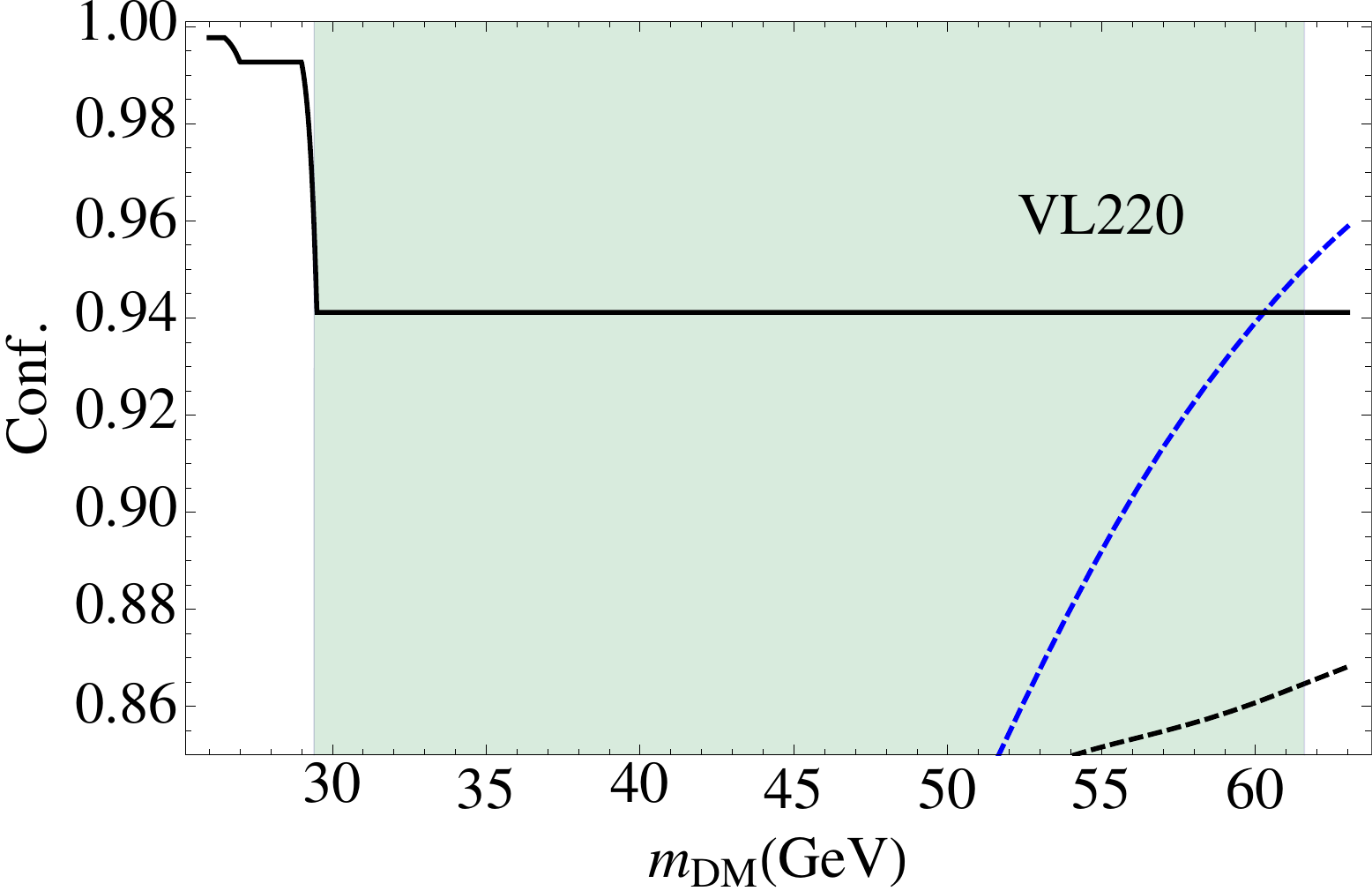}
\includegraphics[width=0.48\textwidth]{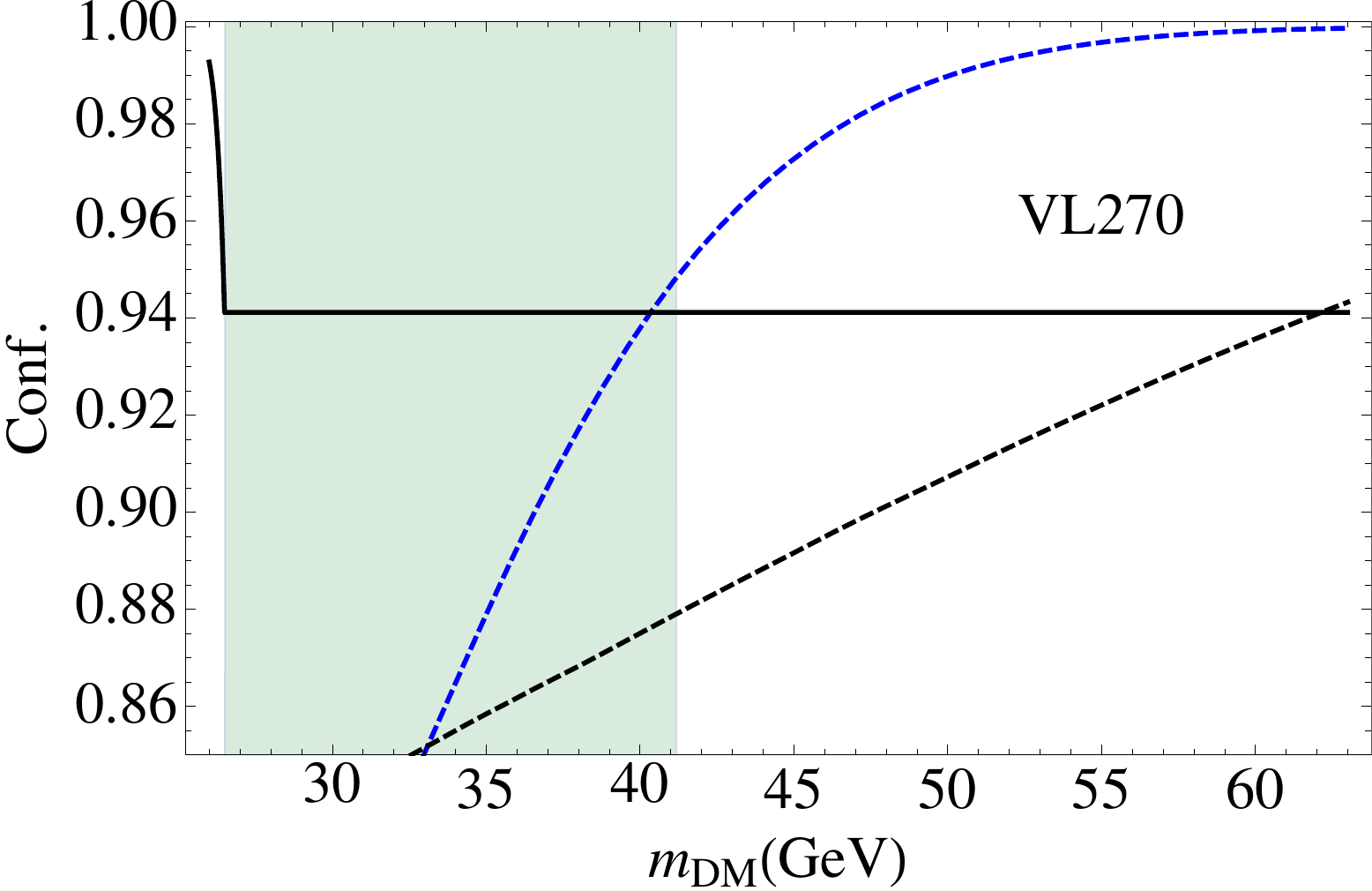}\hspace{2cm}
\includegraphics[width=0.35\textwidth]{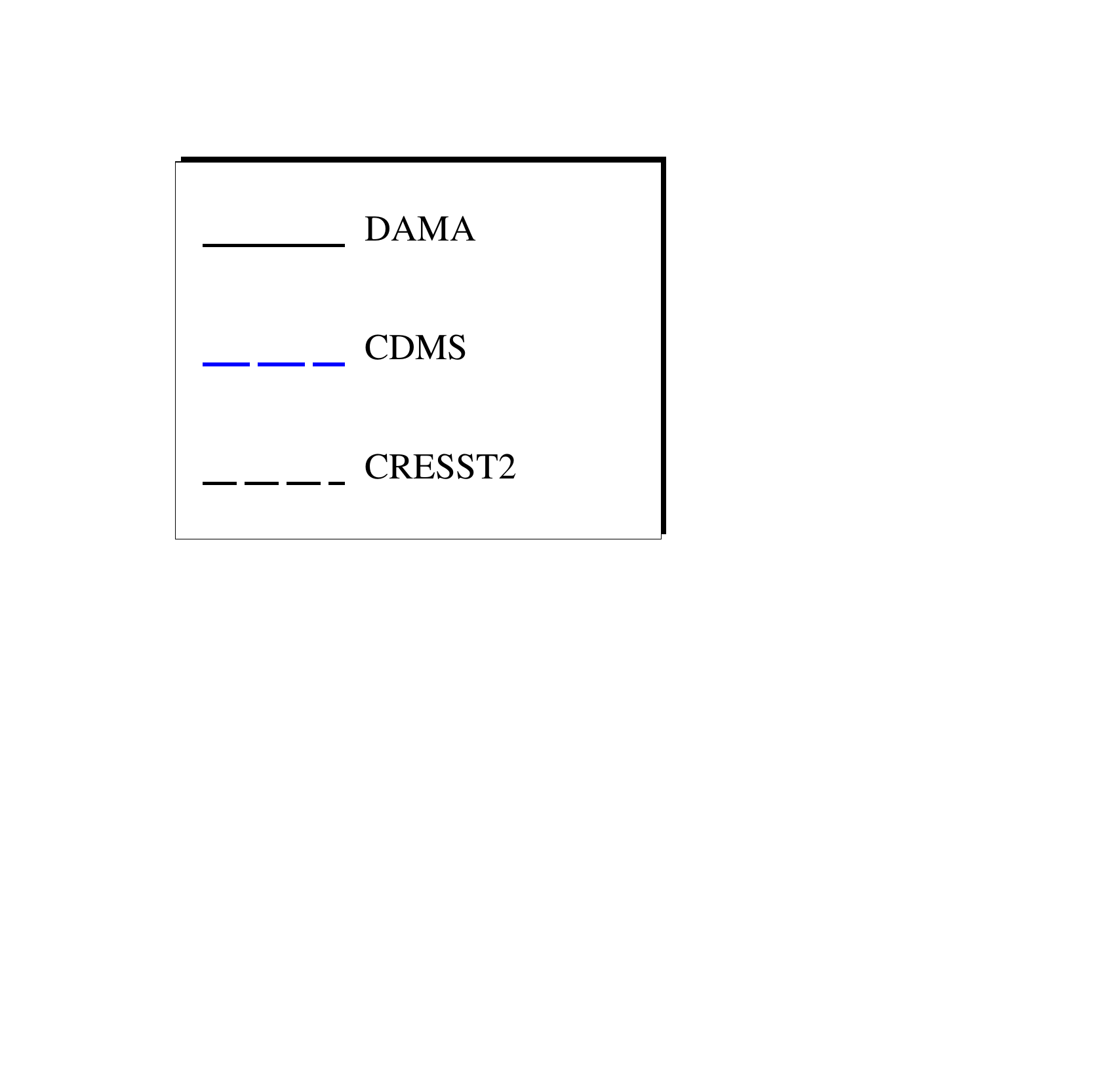}
\caption{Constraints on idealized form factors for different
halo models. For each dark matter mass, we choose a form factor to explicitly
match the
DAMA spectrum, as described in the text.  The confidence levels at DAMA, CDMS
and CRESST2 are then plotted.  We use a $\chi^2$ goodness-of-fit test to
define the confidence at DAMA, and the $pmax$ method for CDMS and CRESST2.  The
standard halo model leaves very little room for a form factor explanation for all of the data.
Highlighted regions have all constraints below 95\%.}
\label{fig:bestformfactor}
\end{center}
\end{figure}

The results of our idealized form factor construction are shown for these
three different halo models in Fig. \ref{fig:bestformfactor}. One can
see that, even with our fairly conservative constraints, there is hardly
any region of masses allowed within the standard halo model.  The
rapid rises in $\chi^2$ for the fit to DAMA occur when the minimum
velocity in an energy bin starts to exceed the bound from the
escape velocity, at which point no form factor is capable of reproducing
the DAMA signal in that bin.
%\footnote{Some discontinuity occurs
%in the null constraints as well at light dark matter masses 
%as an artifact of the unphysical nature of
%our idealized form factor. As energies exceed the kinematically
%accessible range, the form factor stops trying to overcome
%the steep kinematic suppression and we set it to zero; this reduces
%the number of events seen at the null experiments.}
  This suggests
that a purely dynamical explanation of DAMA is disfavored in the standard
halo model\footnote{
The range of allowed masses is somewhat sensitive to the value of
the quenching factor $q_{\rm I}$ for iodine recoils at DAMA, as it
affects the reconstructed nuclear recoil energy range of DAMA's signal.
Here we have taken $q_{\rm I} =0.085$, but if future improved measurements
find it to be lower, then that would move the DAMA spectrum up in recoil energy,
thereby weakening the CDMS constraint and strengthening the CRESST constraint.
For example, in the standard halo model, $q_{\rm I}=0.08$
opens up the 95\% allowed mass range to $30.5 {\rm GeV} 
\le m_{\rm DM} \le 35.5$ GeV. 
However, at $q_{\rm I}=0.07$, the CRESST constraint is stronger than CDMS
and the mass range shrinks to $33 {\rm GeV} \le m_{\rm DM} \le 35$ GeV.
}.  However, there is significantly more room to work with
in the VL$_{270}$ model and, especially, the VL$_{220}$ model\footnote{
We have checked that scattering off of sodium does not work in
any of the three halo models we consider for any range of dark matter masses.}.

At this point, it is worth re-emphasizing some of the main differences between
the iDM mechanism and a dark matter form factor.  
Because the effect of the splitting $\delta$ in eq. (\ref{eq:vmin}) grows
at smaller $q$, one obtains a very rapid kinematic shut-down at low DAMA 
energies,
roughly of the form $\sim e^{-\left( \delta/q\bar{v} \right)^2}$.  
One of the difficulties
of model-building a form factor will be to obtain a sufficiently
rapid shut-down purely from the dynamics.  A more drastic difference
is that $v_{\rm min}$ is a monotonically increasing function of $q$
for elastic dark matter but not for inelastic.  This makes it generally
impossible given the recoil energies and nuclear masses at DAMA and
CDMS to choose parameters so that there are \emph{no} events at
CDMS while simultaneously maintaining the DAMA signal.  However, as 
has been noted in the literature, this is a possibility for inelastic dark matter if
the splitting $\delta$ is chosen sufficiently carefully. We can sharpen
this prediction somewhat by considering the minimum number of events
predicted at CDMS by our idealized form factor in the 95\% goodness-of-fit 
DAMA region (i.e. $\chi^2_{\rm DAMA} < 18.3$).  This 
number is proportional to the effective exposure $\epsilon_{\rm CDMS}$ 
at future CDMS runs, so that 
\be
N_{\rm CDMS, min} = r_{\rm CDMS, min}
{  \epsilon_{\rm CDMS} \over 200 {\rm kg \ day} },
\ee
 where we find that
$r_{\rm CDMS, min} \approx 4.4, 1.1$, and 2.2 for the standard halo,
VL$_{220}$, and VL$_{270}$ models, respectively.

\section{Models}
\label{sec:models}
Here we present models leading to a form factor which is consistent
with present data. One very simple possible form factor, 
which is relatively generic at sufficiently low energies, 
is $F(q)\propto q^2$.  
Such an interaction results in a factor of only $q^4$ in the event rate, and leads to 
insufficient suppression of events at null experiments.  We are thus
lead to introduce a mechanism to further reduce events outside the DAMA region.

Our basic idea is to have multiple gauge bosons
coupling regular matter to dark matter which then naturally  interfere with
one another.   In particular, we will examine two models.  The first  
describes two gauge bosons and leads to an effective form factor

\begin{equation}
\label{2GB}
F_{2GB}(q^2)  \sim  \left( \frac{g_1^2 q^2}{q^2 + m_1^2} - \frac{g_2^2 q^2}{q^2 + m_2^2}\right).
\end{equation}
The second model includes three gauge bosons, leading to 
\begin{equation}
\label{3GB}
F_{3GB}(q^2)  \sim \left( \frac{g_1^2q^2}{q^2 + m_1^2} - 2 \frac{g_2^2  q^2}{q^2 + m_2^2} + 
\frac{g_3^2  q^2}{q^2 + m_3^2}\right).
\end{equation}

In both cases, we will see that the models possess natural regions in parameter space where there is a
custodial symmetry relating the gauge boson masses.  In those regions, the analysis of the models
simplifies.  Specifically, choosing all $m_i$ to be near a GeV, at momenta of interest ($q \ll m_i$),
\begin{eqnarray}
F_{2GB}(q^2)  &=& c_2 q^2 ( q^2 - q_0^2) \\ \nonumber
F_{3GB}(q^2)  &=& c_3 q^2 (q^2 - q_1^2) (q^2-q_2^2).
\end{eqnarray}
Here, the parameters $q_i$ are of order 10 MeV, while the $c_i$ are absorbed into the 
overall crossection.  Obtaining a crossection of the appropriate size in both cases requires $\Lambda, m_i \sim$ GeV.
Therefore, as far as the direct detection experiments are concerned, the two and three gauge boson models contain
two and three parameters, respectively.

\subsection{A two gauge boson model}

Consider two massive gauge bosons $A^{(1,2)}_\mu$ which mix with hypercharge with opposite signs.
\begin{equation}
{\it \L} = \epsilon \left(g_1 F_{\mu\nu}^{(1)} - g_2 F_{\mu\nu}^{(2)}\right) B^{\mu\nu}
\end{equation}
Such a mixing can arise, for example, from integrating out two heavy fields each of which couples
to one of the gauge bosons and to hypercharge with charges chosen 
appropriately.
%\footnote{  The heavy fields should be much lighter than the 
%scale at which the
%mixing first appears, in order for the contribution to be dominated by
%the running, and their masses should not differ from each other by more than 
%an $\CO(1)$ factor, in order for the amount of running to be nearly the same 
%for both terms.}.  

We assume that the dark matter particle is not charged under the gauge bosons, but
that it may be made of constituents that are charged.  Consequently, it develops higher-dimensional
interactions with both gauge bosons. If it is a scalar, the
leading higher-dimensional interactions take the form

\begin{equation}
{\it \L} = i \left(\frac{g_1}{\Lambda^2}  F_{\mu\nu}^{(1)} 
+ \frac{g_2}{\Lambda^2}  F_{\mu\nu}^{(2)}\right) 
\d_\mu X^* \d_\nu X .
\end{equation}
Here $X$ is the dark matter particle and $\Lambda$ is characteristic of its size.  
Thus, the exchange of the two gauge bosons between the dark matter particle and the protons in nuclei will lead
to the form factor of Eq.(\ref{2GB}).

Let us now discuss how the gauge bosons obtain their masses.  As mentioned, the form factor simplifies in the limit that
a custodial symmetry relating the gauge boson masses is weakly broken.  We will therefore describe the Higgs fields and
their potential with that in mind.  Consider a complex doublet, $\Phi = (\phi_1,\phi_2)$.  The custodial symmetry is 
an $SO(2)$ rotation of $\phi_1$ into $\phi_2$.  $\Phi$ couples to the gauge bosons as
\begin{equation}
D_\mu \Phi = \left(\partial_\mu - i \frac{g_1}{2} A^{(1)}_\mu - i \frac{g_2}{2} A^{(2)}_\mu \sigma_3\right) \Phi.
\end{equation}
We now turn to the Higgs potential.  Let us assume in the following that the only source of breaking of the $SU(2)$ symmetry acting on $\Phi$ comes from interactions with $A^{(2)}_\mu$, and that the dark sector interactions respect a charge conjugation symmetry (exchanging $\phi_1$ and $\phi_2$) which takes $A^{(2)}_\mu \rightarrow -A^{(2)}_\mu$.  With these symmetries, the $\Phi$ potential takes the form
\begin{equation}
V(\Phi) = -m^2 \Phi^\dagger \Phi + \lambda (\Phi^\dagger \Phi)^2 + y (\Phi^\dagger  \sigma_3 \Phi)^2. 
 \end{equation}
If $y>1$, at the minimum $\Phi^\dagger  \sigma_3 \Phi = 0$.  This ensures that the gauge boson mass
matrix is diagonal and that there is an $SO(2)$ custodial symmetry relating the gauge boson masses, such that $m_{1,2}= g_{1,2} ~v$. 
Including terms in the potential which violate the charge conjugation symmetry would make this potential generic, resulting in
general gauge boson masses.  For our purposes, however, it is useful to assume the there is only a small violation 
of the above custodial symmetry, such that $\frac{m_1^2}{g_1^2} - \frac{m_2^2}{g_2^2} = (\delta m)^2$.
Such a violation could come, for instance, from the above mixing of the gauge bosons with  hypercharge (which violates the 
charge conjugation symmetry at order $\epsilon^2$).  In this case, the size of $(\delta m)^2$ would depend on additional physics
such as the cutoff scale of the $\Phi$ theory, $(\delta m)^2 \sim \frac{\epsilon^2}{16\pi^2} \Lambda_{\rm cutoff}^2$.  
One can imagine other sources for $(\delta m)^2$ as well, and we therefore leave it as a free parameter.
Thus, in the limit of interest, where $q^2  \ll m_{1,2}^2$, the form factor is approximated by
\begin{equation}
F_{2GB}(q^2)  \sim  \frac{q^2}{v^4} \left( \frac{g_1^2-g_2^2}{g_1^2 g_2^2}~ q^2 - (\delta m)^2 \right) \equiv c_2 q^2 (q^2 -q_0^2).
\end{equation}
We note that besides the constant $c$ that can be reabsorbed into the total crossection, the only relevant parameter for 
dark matter detection experiments is $q_0$.

\subsection{A three gauge boson model}

One can also consider extending the above scenario to include an additional gauge boson which mixes with hypercharge.  The interference between the three gauge bosons can naturally achieve further suppression of the cross-section at low momentum transfer.  The resulting model is similar to the model above, but with an $SO(3)$ custodial symmetry instead.  As before, 
the three gauge bosons mix with hypercharge,
\begin{equation}
{\it \L} = \epsilon \left(g_1 F_{\mu\nu}^{(1)} - 2 g_2 F_{\mu\nu}^{(2)} + g_3 F_{\mu\nu}^{(3)} \right) B^{\mu\nu}.
\end{equation}
Such a mixing can be obtained by integrating out three heavy fields, each charged under hypercharge and 
under only one of gauge bosons.  All heavy fields have the same charge under their respective gauge bosons,
and charges $+1$, $-2$, and $+1$ under hypercharge.  

The dark matter particle couples to the gauge bosons through 
the leading higher-dimensional interactions,
\begin{equation}
{\it \L} = i \left(\frac{g_1}{\Lambda^2}  F_{\mu\nu}^{(1)} + \frac{g_2}{\Lambda^2} F_{\mu\nu}^{(2)} + \frac{g_3}{\Lambda^2} F_{\mu\nu}^{(3)}\right) 
\d_\mu X^* \d_\nu X .
\end{equation}
Hence, gauge boson exchange results in the effective form factor of eq.(\ref{3GB}).  

We will again concentrate on a version of this model with approximate custodial symmetry.  To that end, 
we take the Higgs field, $\Phi$, to be a four of $SU(4)$, charged under the gauge bosons as
\begin{equation}
D_\mu \Phi = \left(\d_\mu - i g_i A^{(i)}_\mu Q_i\right) \Phi.
\end{equation}
Here, the charges, $Q_i$ are embedded in $SU(4)$ as
\begin{equation}
Q_1 =
\begin{pmatrix}
\openone& \quad \\ 
\quad & \openone   
 \end{pmatrix},
 Q_2 = 
\begin{pmatrix}
\openone& \quad \\ 
\quad & -\openone   
 \end{pmatrix},
 Q_3 = 
\begin{pmatrix}
\sigma_3 & \quad \\ 
\quad &   \sigma_3
 \end{pmatrix},
\end{equation}
where $\openone$ is the two by two identity matrix.  Assuming that the only interactions breaking the $SU(4)$ symmetry are the gauge interactions, a general potential respecting all manner of charge conjugation symmetries is 
\begin{equation}
V(\Phi) = -m^2 \Phi^\dagger \Phi + y_i (\Phi^\dagger  Q_i \Phi)^2 + 
\tilde{y} (\Phi^\dagger  Q_2 Q_3 \Phi)^2. 
 \end{equation}
If $y_i, \tilde{y} > 0$, then at the minimum an $SO(3)$ custodial symmetry acting on the gauge boson masses will be 
respected, and $m_i = g_i v$.  This custodial symmetry will be violated at order $\epsilon^2$.  We thus define
$(\delta m)^2=\frac{2m_2^2}{g_2^2} - \frac{m_1^2}{g_1^2} - \frac{m_3^2}{g_3^2}$.

In order to obtain a simplified version of this model at low $q$, we will further require that the three gauge bosons unify at some 
scale $M_U$.  For instance, the unified group can be $SO(8) \supset SU(4) \times U(1)$, with $\Phi$ transforming as the eight
of $SO(8)$.  Therefore, $g_i(M_U) = g_U$.  Below $M_U$, we assume that the matter content is such that
the running splits the couplings, but maintains the relation $\frac{1}{g_1^2} -\frac{2}{g_2^2} + \frac{1}{g_3^2}=0$.
This is possible if for example there are twice as many fermions charged under only $U(1)_1$ as under only $U(1)_2$,
with no new matter (besides $\Phi$) charged under $U(1)_3$.  Threshold effects would then account for a violation of 
the relation between couplings, so that $\frac{1}{g_1^2} -\frac{2}{g_2^2} + \frac{1}{g_3^2} \equiv \delta g \lesssim \frac{1}{100}$.
Consequently, at $q \ll m_i$,
\begin{equation}
F_{3GB}(q^2) \sim  \frac{q^2}{v^6} \left[\left(\frac{1}{g_1^4} -\frac{2}{g_2^4} + \frac{1}{g_3^4}\right) q^4 - \delta g ~v^2 q^2 + (\delta m)^2 v^2\right]\equiv c_3 q^2 (q^2 - q_1^2) (q^2-q_2^2).
\end{equation}
We note that choosing $\Lambda_{\rm cutoff} \sim 4\pi v$ naturally gives $(\delta m)^2 v^2 \sim 10^{-6} v^4$, which is 
the appropriate size for $q_i^4$ (assuming all couplings are of order one).

\section{Null Experiments and Analysis}
\label{sec:null}

We will briefly describe the null experiments whose constraints we
include and the assumptions that go into our analysis.  We have attempted
to follow the analysis assumptions 
of \cite{SpencerNeal} in order to make the comparison
with inelastic dark matter transparent.  Summary tables of the assumptions
at each experiment are \ref{tab:dama},\ref{tab:nullexpts}, and \ref{tab:events}.
 Tables \ref{tab:nullexpts} and \ref{tab:events} show the relevant data used in
our analysis concerning the various null experiments.  
Generally, these experiments see some set of
unexplained events whose energies are indicated in table \ref{tab:events}, 
and in our analysis we have
allowed ourselves to interpret these as potential dark-matter-induced 
nuclear recoils.  
The energies listed in this table are actual nuclear recoil energies, and have
thus been obtained by dividing the detected energy by a quenching factor
where appropriate.
In all null experiments in which individual potential events are 
published, we calculate the constraints using the $pmax$, or
``maximum gap'',  value
\cite{Yellin}.  This is the probability, given a specific model,
 of an experiment having seen no events in the gap between any two adjacent 
potential events.
When the number of observed events is very low, this method is
very sensitive to, for
example, non-blind cuts on events, since a single additional
event in the middle of a gap can drastically increase the likelihood.

\begin{table}[t]
  \begin{tabular}{|l|l|}
    \hline
    Bin Energy &  Modulating Amplitude \\
    (keVee)   & (cpd/kg/keVee) \\
    \hline \hline
    $2.25 \pm .25$ & $.015 \pm .0045$\\
    \hline
    $2.75 \pm .25$ & $.027 \pm .0050$\\
    \hline
    $3.25 \pm .25$ & $.023 \pm .0050$\\
    \hline
    $3.75 \pm .25$ & $.013 \pm .0048$\\
    \hline
    $4.25 \pm .25$ & $.0093 \pm .0043$\\
    \hline
    $4.75 \pm .25$ & $.011 \pm .0038$\\
    \hline
    $5.25 \pm .25$ & $.0035 \pm .0038$\\
    \hline
    $5.75 \pm .25$ & $.0025 \pm .0038$\\
    \hline
    $6.25 \pm .25$ & $.0055 \pm .0035$\\
    \hline
    $6.75 \pm .25$ & $.0013 \pm .0033$\\
    \hline
    $7.25 \pm .25$ & $.00025 \pm .0035$\\
    \hline
    $7.75 \pm .25$ & $.00025 \pm .0035$\\
    \hline
  \end{tabular}
 \begin{tabular}{|l|l|}
    \hline
    Bin Energy & Amplitude \\
    (keVee)   & (cpd/kg/keVee) \\
    \hline \hline
    $3.5 \pm 1. $ & $.11 \pm .099$\\
    \hline
    $4.5 \pm 1.$ & $.11 \pm .078$\\
    \hline
    $5.5 \pm 1.$ & $.027 \pm .069$\\
    \hline
    $6.5 \pm 1.$ & $.074 \pm .054$\\
    \hline
    $7.5 \pm 1.$ & $-.0008 \pm .037$\\
    \hline
    $8.5 \pm 1.$ & $.051 \pm .036$\\
    \hline
    $9.5 \pm 1.$ & $.051 \pm .027$\\
    \hline
    $10.5 \pm 1.$ & $.065 \pm .026$\\
    \hline  
  \end{tabular}
  \caption{Left: amplitude of the modulating part of the DAMA signal, 
divided into 0.5keVee bins. Right: amplitude of the measured rate
at KIMS, in 1.0 keVee bins. }
\label{tab:dama}
\end{table}

\begin{table}[t]
  \begin{tabular}{|l|l|l|l|l|}
    \hline
    Experiment & Element & Dates & Effective Exposure & Signal
    Window\\
   & & & (kg days) & (keV) \\
    \hline \hline
    CDMS & Ge & Oct. 2- July 31 & 174.4 & 10 - 100\\
    \hline
    XENON & Xe & Oct. 1 - Feb. 28 & 122$^*$ & 2 - 70\\
    \hline
    CRESST-II, '04  & CaWO$_4$ & Jan. 31 - Mar. 23 & 18.5 & 12 - 50\\
    \hline
    CRESST-II, '07  & CaWO$_4$ & Mar. 27 - July 23 & 43.1 & 12 - 100\\
    \hline
    ZEPLIN-II & Xe & May 1 - June 30 & 112.5 & 13.9 - 55.6\\
    \hline
    ZEPLIN-III & Xe & Feb. 27 - May 20 & 126.7 & 10.7 - 30.2\\
    \hline
    KIMS & CsI & see text   & N/A & 20.6 - 65.4\\
    \hline
  \end{tabular}
  \caption{Elements, dates, effective exposures and signal windows for the
    null experiments.  The exposures have been listed here only after
    efficiencies and cuts have been taken into account (it is still
    necessary, however, to multiply by the fraction of target nuclei in the
    detector material). $^*$XENON effective exposure is approximate,
    since the efficiencies and acceptance rates are energy-dependent.
    This was taken into account in our analysis.}
\label{tab:nullexpts}
\end{table}

\begin{table}[t]
  \begin{tabular}{|l|l|}
    \hline
    Experiment & Events (keV)\\
    \hline \hline
    CDMS & 10.5, 64\\
    \hline
    XENON & 15.8, 23.8, 24.1, 25.9, 32.2, 33.7, 38.7, 39.2, 40.1, 43.3, 43.6, 
        51.6, 62.3 \\
    \hline
    CRESST-II '04 & 13, 18.5, 22.5, 24, 33.5\\
    \hline
    CRESST-II '07 & 16.7, 18.03, 33.09, 43, 43, 43, 63\\
    \hline
    ZEPLIN-II & 13.9, 15.3,16.7,16.7,16.7,18.1,\\
  &   19.5,19.5,20.9,20.9,22.2,22.2, \\
  &    23.6,27.8,27.8, 29.2,30.6, 33.4,\\
  &  37.5, 37.5,  38.9, 41.7,  43.1, 44.5, \\
  &  44.5, 48.7, 50.0, 50.0, 54.2 \\
    \hline
    ZEPLIN-III & 14.6, 17.2, 20.8, 22.9, 23.4, 25.1, 28.1\\
    \hline
  \end{tabular}
  \caption{Unexplained events observed by the null experiments, listed by
    nuclear recoil energy in keV.}
\label{tab:events}
\end{table}

\subsection{DAMA}

We take for the DAMA signal the published modulation amplitudes from
Fig. 9 of \cite{DAMALIBRA}, which we summarize in Table \ref{tab:dama}.
The energy shown
is that observed in the detector, and is smaller than
the actual nuclear recoil energy by an amount given by the ``quenching
factor''. For Iodine scattering events in a NaI crystal, we take
the  quenching factor to be 0.085. 
We implement scattering off Iodine only, as our form factor suppresses
events scattering off of Sodium.  We calculate the modulation amplitude
as half of the difference between the event rate on June 2 and Dec. 2.  
To obtain our confidence limits on our models from DAMA,
we find the best $\chi^2$ at a fixed mass and show contours of
$\chi^2 = \chi^2_{\rm min} + 4.61 (9.21)$ for 90\% (99\%) confidence.

\subsection{CDMS}

We combine the CDMS data from the three runs in \cite{CDMS, CDMS2, CDMSFIVE},
for a total of 174.4 kg days after including their published 30\%
acceptance rates on a standard WIMP.  The first run had 20 kg-days and
ran from Oct. 11 - Jan 11, the second had 34 kg-days and ran from
Mar. 25-Aug 8, and the latest five-tower run (which saw no events)
had 121 kg-days and
ran from Oct. to July.  

\subsection{CRESST-II}

We combine the CRESST-II data from the 2004 run \cite{CRESST2004}
and the latest commissioning run \cite{CRESST2} from 2007. The
'04 run ran from Jan. 31 - Mar. 23, had 20.5 kg-days of
exposure, and looked at recoil energies of 12-50 keV.  The
'07 run ran from Mar. 27 - July 23, had 47.9 kg-days of exposure,
and looked at 12-100 keV. Both runs had very mild cuts leading
to a 90\% acceptance rate.  The '04 run had two detectors, one of
which (JULIA) saw four events, and the other of which (DAISY) saw
no events using the published quenching factors, and one
events after varying the quenching factors within their published
uncertainties.  The '07 run also had two detectors, which saw a total
of seven events.  We should note that the $pmax$ constraints from the '07 run 
alone are much stronger than the constraints from the combined '04/'07
analysis, but given that the '07 run was non-blind we follow
\cite{SpencerNeal,Lang:2009ge} and take all the CRESST-II data together.

\subsection{XENON}

For XENON10 \cite{XENON10}, we take the published data of the recent 
reanalysis \cite{XENON10IDM}
 in the extended range 2.0-70 keV.
%\footnote{Here we differ from \cite{SpencerNeal,JMR} in that we
%do not include the unpublished data above 27 keV.}.
The experiment ran from Oct. 1- Feb 28. and had a total of 316 kg-days.
We take the published energy-dependent acceptance rates and efficiencies.

\subsection{ZEPLIN-II and ZEPLIN-III}

ZEPLIN-II is a Xenon based experiment with 225 kg-days of effective 
exposure after
cuts are applied, and in addition there is 
a recoil detection efficiency of 50\%.
We take the quenching factor to be 0.36, and the 29 events from
Fig. 8 of \cite{ZEPLIN2}.  We take all events as potential signal
in calculating our limits.  

ZEPLIN-III ran in 2008 from Feb. 27 - May 20 and observed only
7 events between 10.7-30.2 keV \cite{ZEPLIN3}.

\subsection{KIMS}

KIMS \cite{KIMS} had four CsI(Tl) crystals looking in the range 3-11 keVee,
two of which ran roughly year-round and two of which ran between June and
March. We take the weighted average over all four crystals to 
obtain the measured rate and error, shown in table \ref{tab:dama}.  
 We use the parameterization of the quenching factor from \cite{JMR}:
\be
q_{CsI} (E_R) &=& { 0.175 e^{-E_R/137}  \over 1 + 0.00091 E_R},
\ee
and define the constraint as the requirement 
that the total scattering off Cesium and Iodine not
be more in any of the first five bins (3-8 keVee) than the
measured rate plus 1.64 times the error.

\section{Results}
\label{sec:results}

\begin{figure}[t!]
\begin{center}
\includegraphics[width=0.40\textwidth]{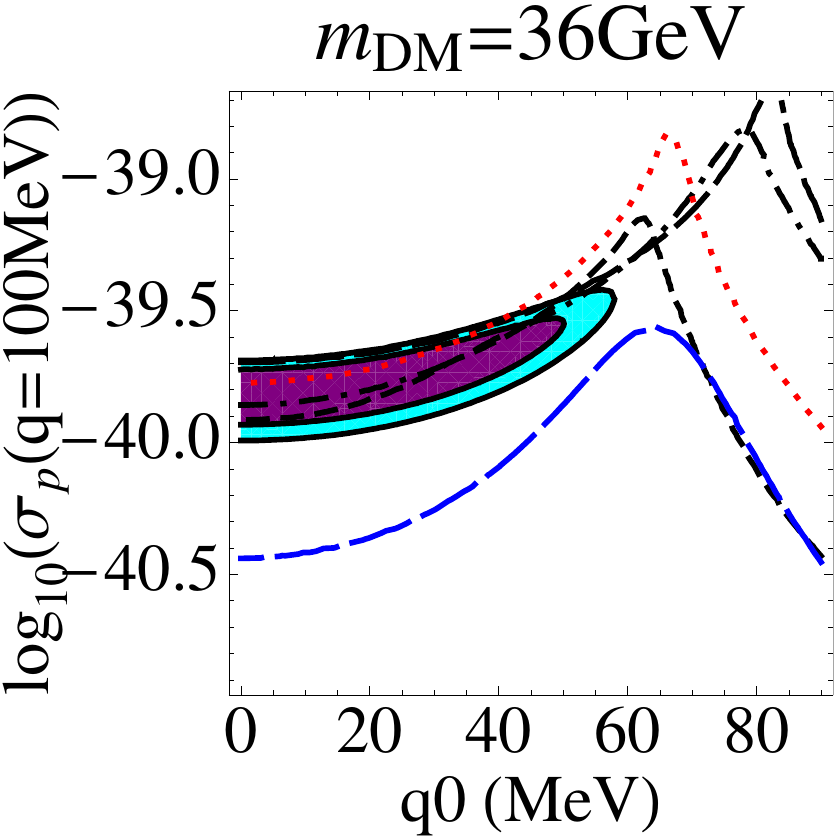}
\includegraphics[width=0.40\textwidth]{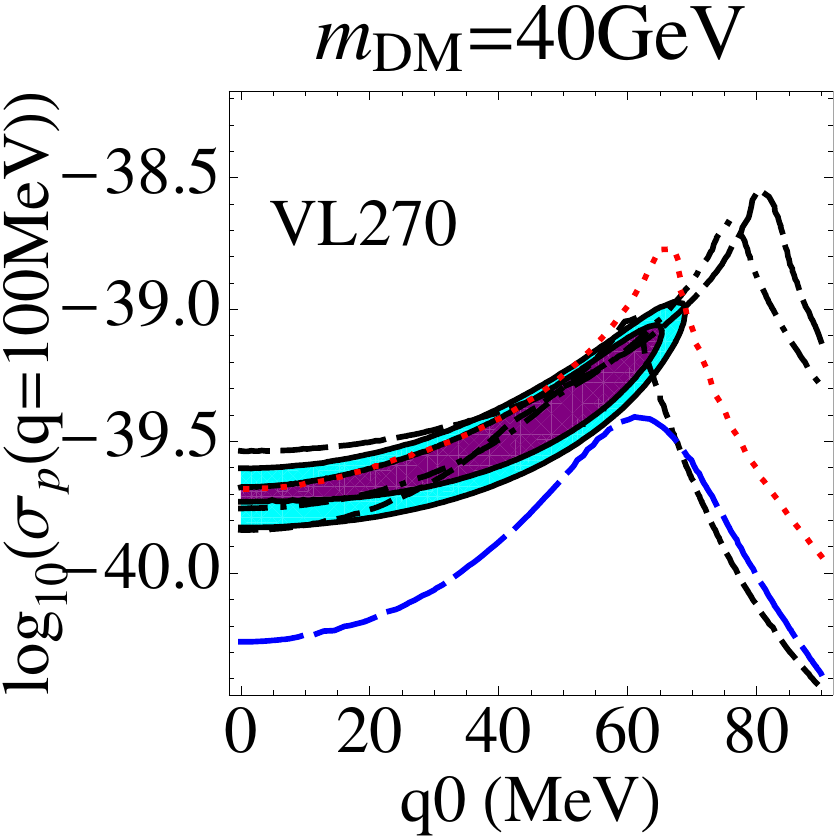}
\includegraphics[width=0.40\textwidth]{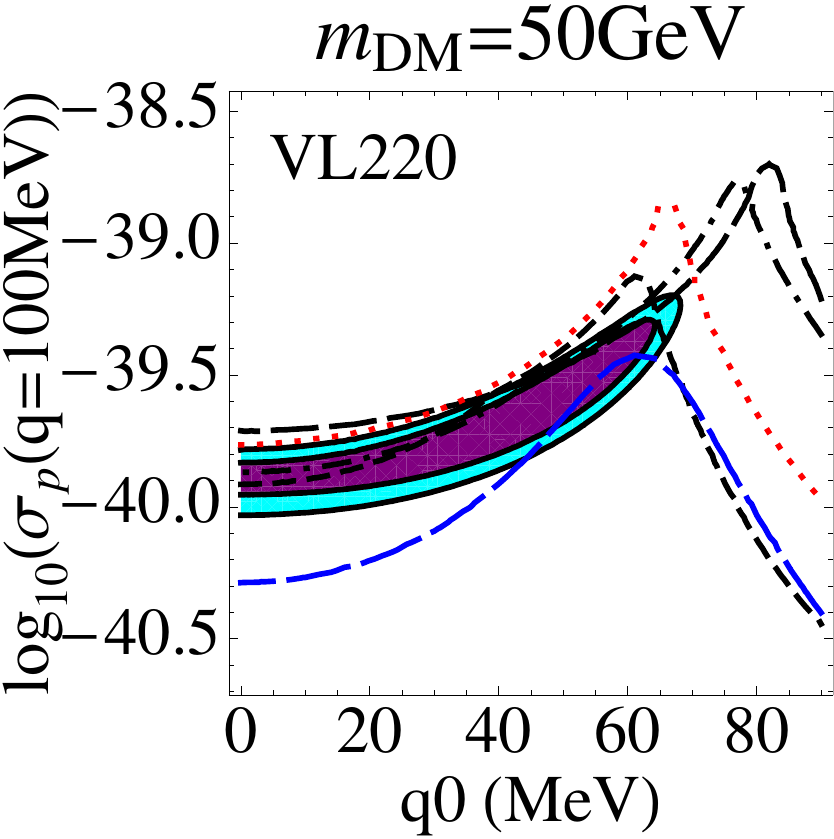}
\includegraphics[width=0.40\textwidth]{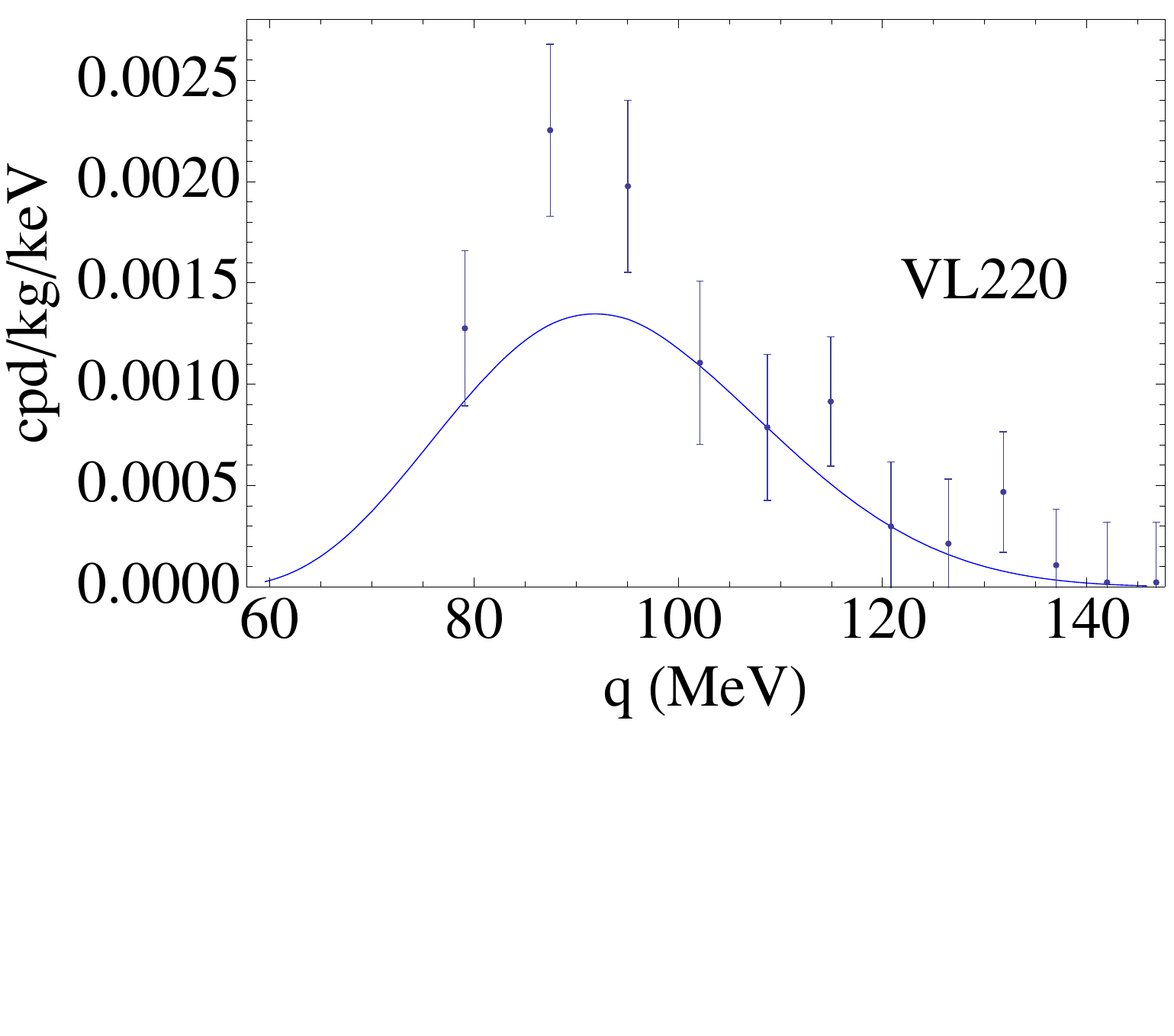}
\includegraphics[width=0.40\textwidth]{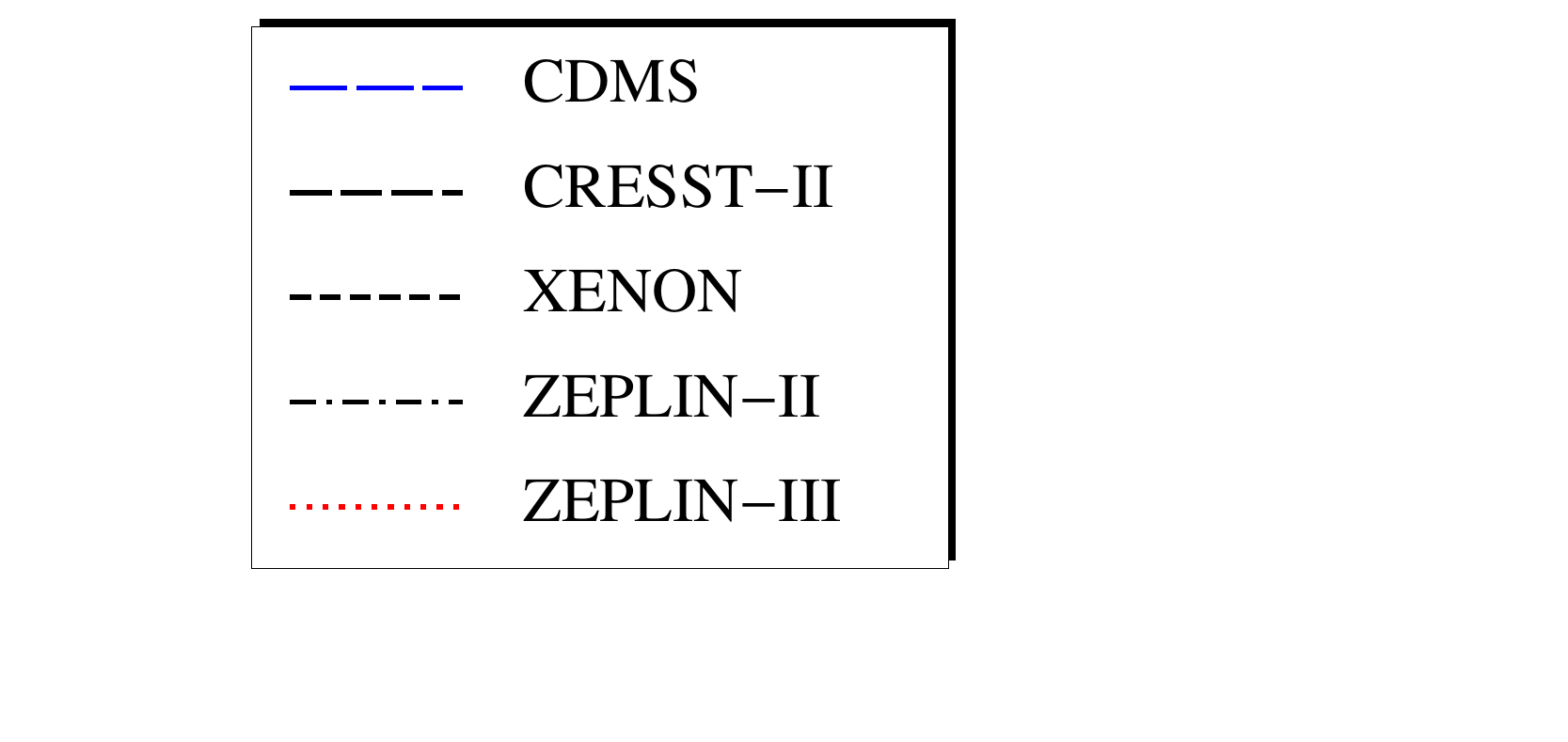}
\caption{ Exclusion plots for the two-gauge-boson (2GB) model. 
Confidence limits from DAMA are shown in purple (light blue)
for 90\% (99\%), all other confidence limits are 99\%. The spectrum
at DAMA is shown for a benchmark point with VL$_{220}$; see table \ref{tab:bench} for the model parameters. $\sigma_p$ is taken in units of cm$^2$.}
\label{fig:GB2contours}
\end{center}
\end{figure}

\begin{figure}[h!]
\begin{center}
\includegraphics[width=0.38\textwidth]{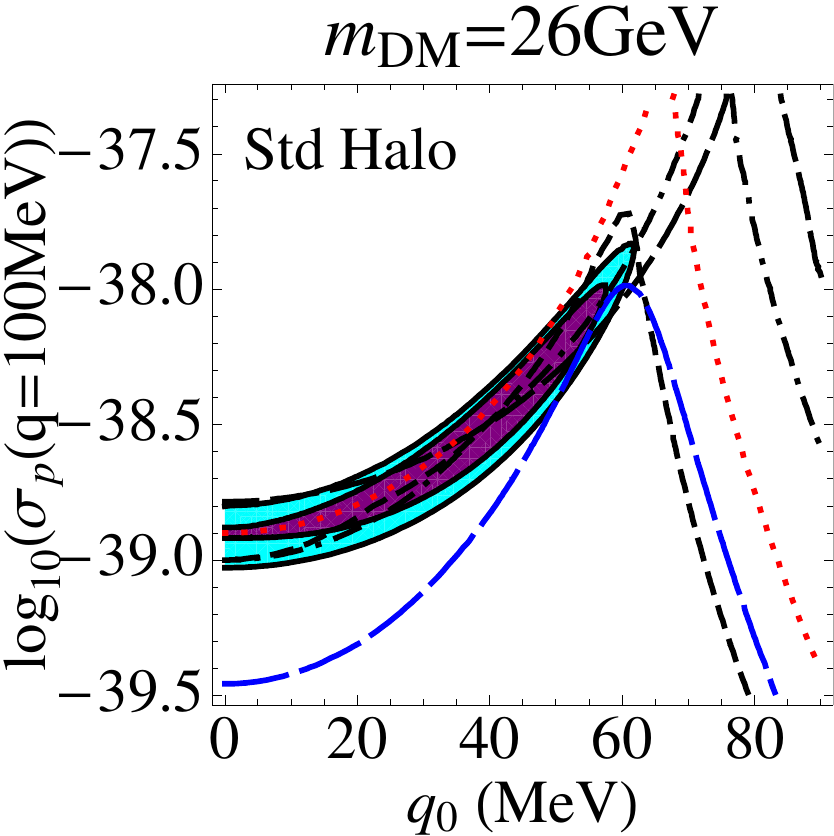}
\includegraphics[width=0.38\textwidth]{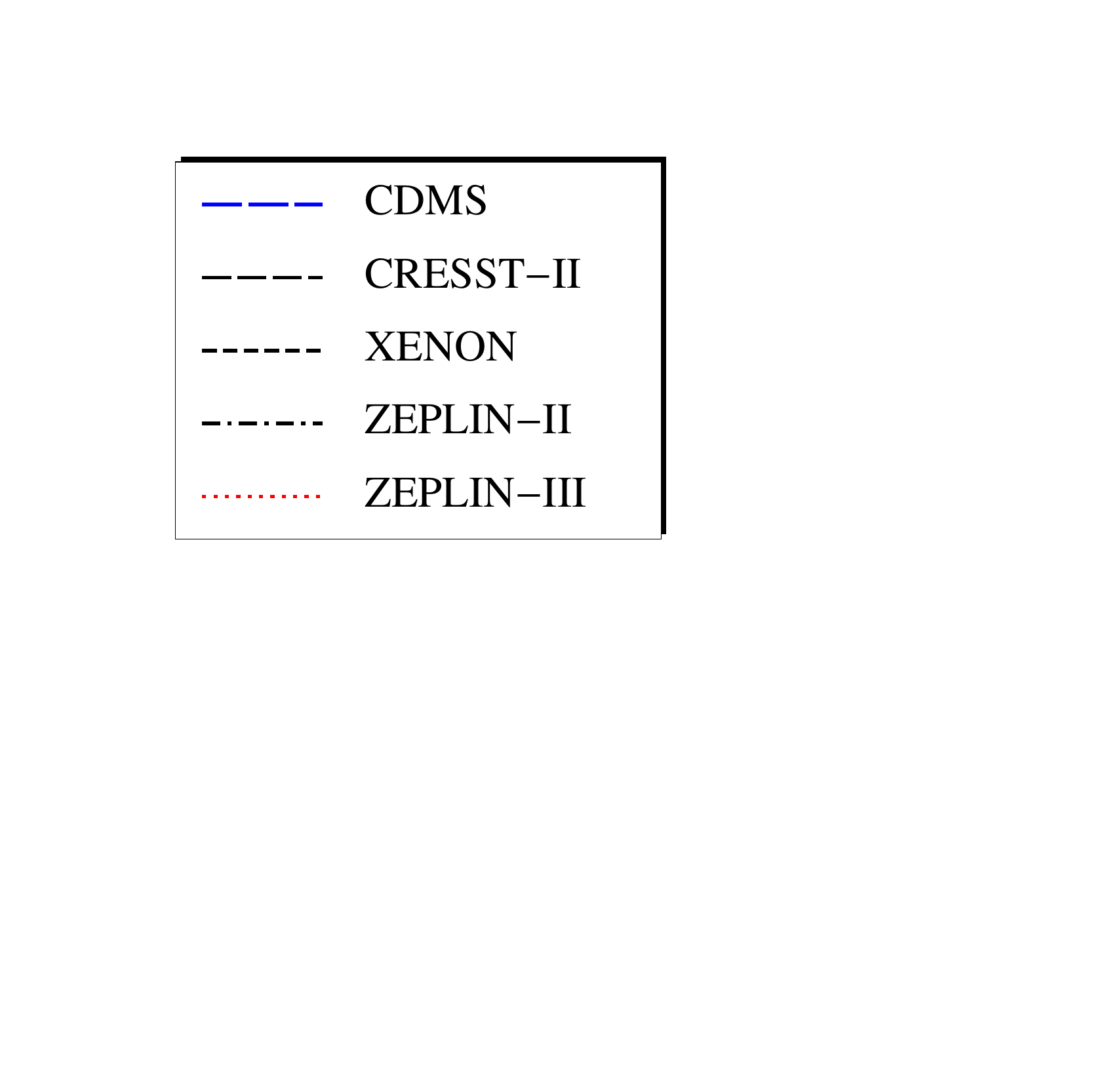}
\includegraphics[width=0.38\textwidth]{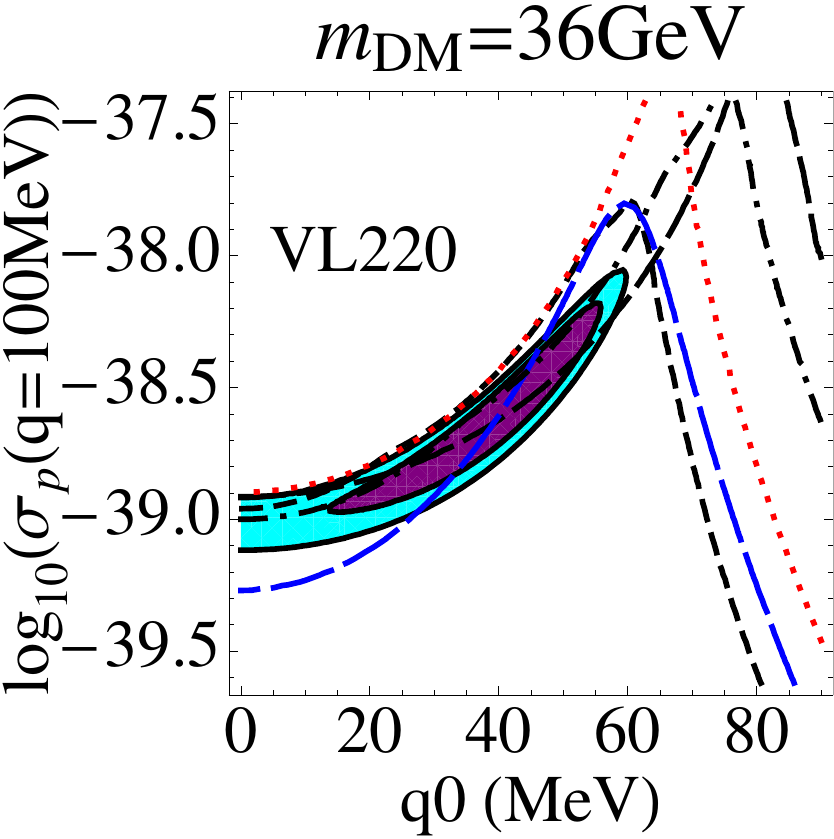}
\includegraphics[width=0.38\textwidth]{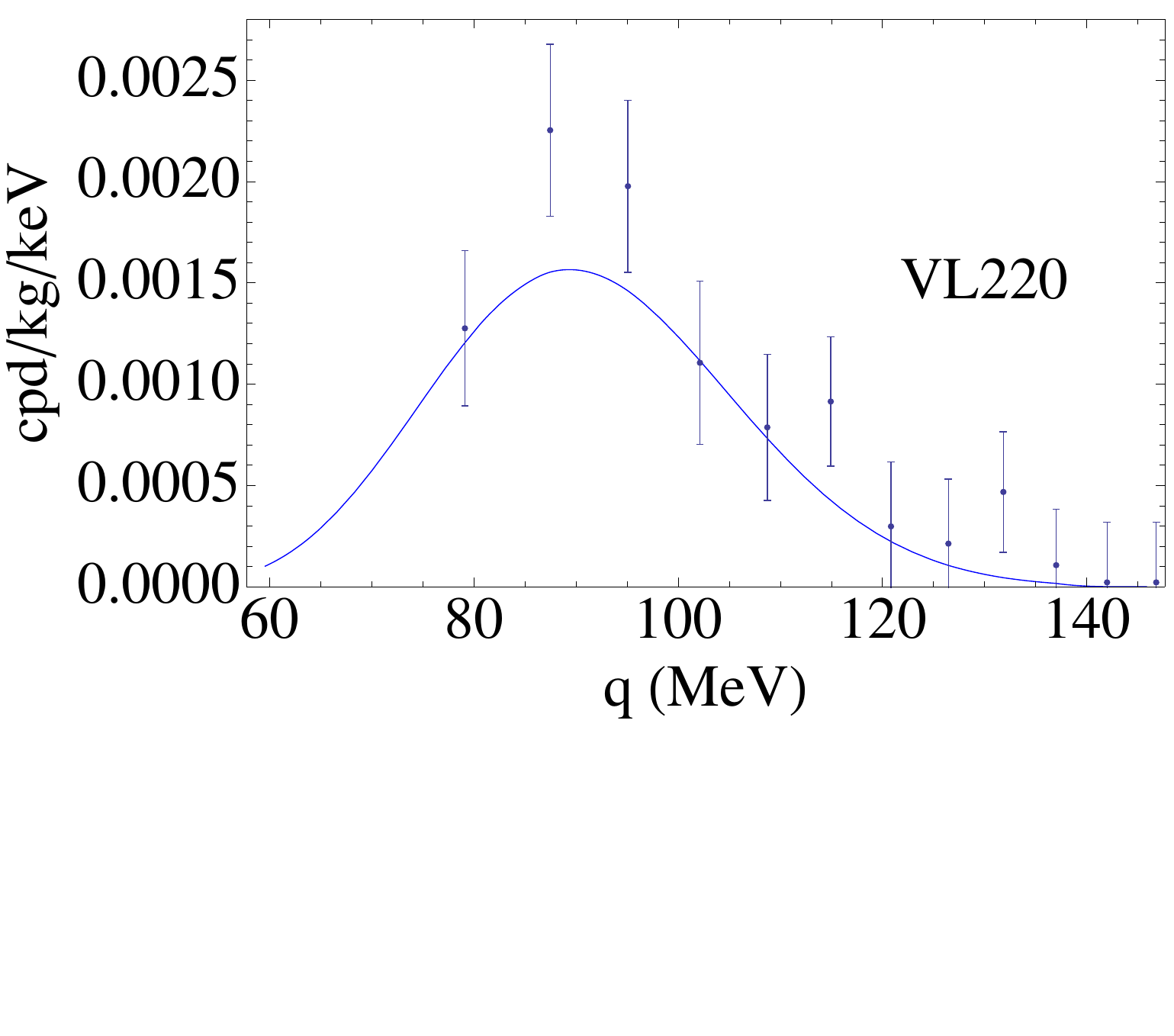}
\includegraphics[width=0.38\textwidth]{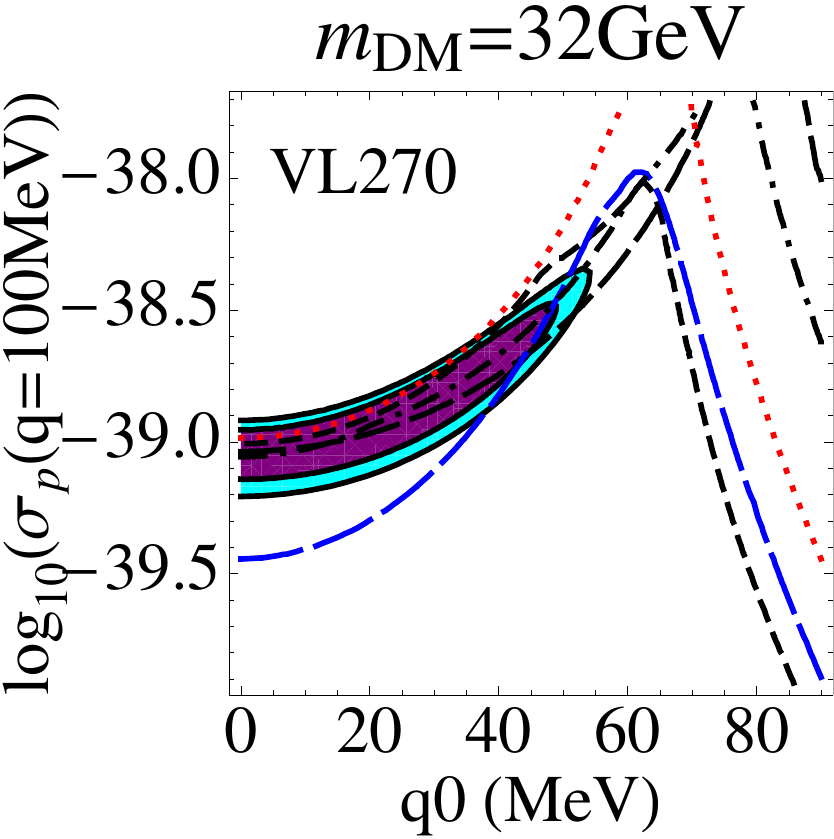}
\includegraphics[width=0.38\textwidth]{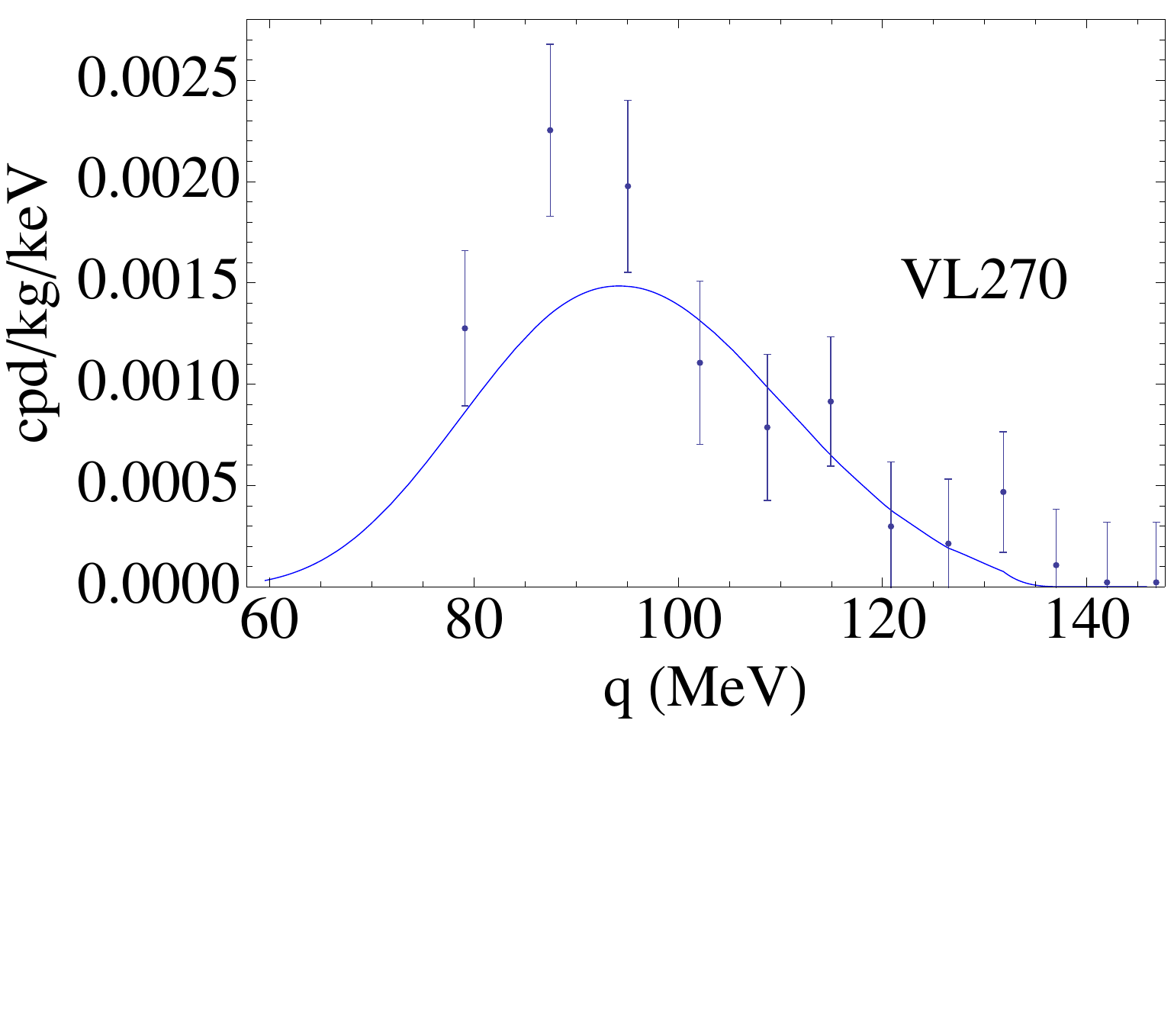}
\caption{ Exclusion plots for the three-gauge-boson model (3GB) with
both nodes taken  at the same value $q_1=q_2\equiv q_0$ for
simplicity. Confidence limits from DAMA are shown in purple (light blue)
for 90\% (99\%), all other confidence limits are 95\%. 
The spectrum
at DAMA is shown for two benchmark points; see table \ref{tab:bench} for the model parameters.  
}
\label{fig:GB3contours}
\end{center}
\end{figure}

Exclusion plots are shown in Figs. \ref{fig:GB2contours} and
\ref{fig:GB3contours}\footnote{
The dark matter-proton cross-section $\sigma_p$ now involves the
form factor, and so is a function of $q$.  We therefore have defined
$\sigma_{p,100 {\rm MeV}}$ in the following way: $\sigma_p =
\sigma_{p,100{\rm MeV}} F_{\rm DM}^2(q)$, where $F_{\rm DM}(q)$
is normalized e.g. as $q^2 (q^2- q_1^2)(q^2-q_2^2)/(100{\rm MeV})^6$.  } 
for the two- and three-gauge boson (2GB and
3GB) models,
respectively.  In order to reduce the parameter space and thus
simplify the contour plots, we have taken both nodes in the
3GB model to be at $q_1=q_2 \equiv q_0$; the improvement
in consistency with the null experiments by varying them
separately is relatively small.
We have checked that in each case, the constraint from KIMS can be neglected 
since it is weaker than both the CDMS and CRESST-II constraints, 
and in fact its 95\% 
constraint curve does not cut out any of the DAMA 90\% preferred 
region\footnote{We note that in our models the constraint from comparing the 
unmodulated DAMA signal to the
modulated DAMA signal is an even weaker constraint on the modulation fraction 
than KIMS is, and so may also be neglected.  The danger of overpredicting the 
unmodulated signal in low energy
bins $E_R < 20 $ keV is avoided as we predict the spectrum to decrease at low 
energies.  }. 
The DAMA inner (outer) contours are shown at 90\% (99\%),
and the $pmax$ constraints for the null experiments are shown
at 99\% (95\%) for the 2GB (3GB) model\footnote{
The reason for this difference is that the DAMA 99\% preferred region
for the 2GB model is just barely excluded at 95\% by our analysis.  
We hope this will not cause
too much confusion.}.  
There is a fairly strong dependence of the limits on the dark matter
mass, both at DAMA and at the null experiments.  As discussed in
section \ref{sec:qoverlap}, sufficiently low masses put the
DAMA signal above $v_{\rm esc}$ and thus are excluded. On the
other hand, sufficiently high masses cause the
velocity dispersion to flatten out at low recoil energies.
In %the context of 
our specific models, the lower energy
bins are too suppressed by the form factor and the spectrum
is incorrectly distorted.  On top of this, larger masses
do not lead to a large enough modulation fraction at DAMA
in order to evade the null constraints. In Figs.
\ref{fig:GB2contours} and \ref{fig:GB3contours}, we have
chosen masses from around the regions where the constraints
are weakest.

\newpage

The DAMA preferred region for the 2GB model is excluded
by our analysis at 95\% confidence.  In all cases, the culprit is CDMS,
due to the fact that the form factor does not fall quite rapidly enough
below the DAMA signal region.  We find though that at VL$_{220}$ the 95\%
CDMS line just barely excludes the DAMA preferred region, and there
are still a number of uncertainties (e.g. in quenching factors, energy
resolution) that we have not attempted to take into account.
The situation is much better in the 3GB model, where we show 95\%
exclusion limits. In the standard halo model, no parameter space is
left at 90\%, but for the Via Lactea models the DAMA preferred region
is just barely allowed by our exclusion limits.  The CDMS constraint
is considerably weaker for the 3GB model, due to the sharper fall-off
of the form factor at energies just below the DAMA signal.  As
one might have expected, the other experiments
with a sizeable range below the DAMA signal (XENON, ZEPLIN-II, and ZEPLIN-III)
have also gotten weaker relative to CRESST-II. One can clearly see the
effect of the node $q_0$ as it passes through the bulk of each individual
experiment, removing the predicted events in that region.  For 
$q_0$ above about 70 MeV (60 MeV) in the 2GB (3GB) model, 
however, the node starts to suppress the DAMA signal, 
whereas ideally the form factor will be more or less
set in this region by the DAMA spectrum.  Around $q_0 \sim 80$ MeV,
the CRESST-II signal is drastically suppressed by the node.
Thus the trend in both models is for the lower-$q$ experiments
to dominate the constraints at smaller $q_0$, whereas for
40 MeV $\lesssim q_0 \lesssim $ 70 MeV, they are suppressed and
CRESST-II becomes the dominant constraint.

The halo model with the most open parameter space is in all cases VL$_{220}$.
The first major effect of the smaller $\bar{v}$ at VL$_{220}$ is to
tighten the velocity distribution.
This allows for larger dark matter masses without sacrificing the 
necessary modulation fraction.  Additionally, larger masses effectively
push the escape velocity cut-off to higher recoil energies at DAMA, improving
the fit to the upper energy bins.  
Finally, it is important to note that despite the tighter distribution
at higher energies, the distribution in the lab frame 
is still relatively flat below the earth's velocity, as the velocity
of such events are 
dominated by the earth's contribution rather than the halo fluctuations.
Given how much a modest change (18.5\%) in $\bar{v}$ can affect the allowed
parameter space, it is clearly an important issue to understand the
experimentally and theoretically allowed range of parameter space for
halo models, so that it would be possible to more systematically include 
all such uncertainties in setting exclusion limits.
In the absence of such an analysis, the precise meaning of
exclusion limits is not entirely clear.

\begin{table}[t]
\begin{center}  \begin{tabular}{|l|l|l|l|c|c|c|}
    \hline
    Halo Model & DM Model & $m_{\rm DM}$ & $\chi^2_{\rm DAMA}$ &
$\sigma_{p,100 {\rm MeV}}$ & $p_{\rm CDMS}$ & $p_{\rm CRESST}$ \\
       \hline \hline
  VL$_{220}$ & 2GB $(q_0 = 50  {\rm MeV})$  & 50 GeV & 11.8  & $2.34 \times 10^{-40} {\rm cm}^2$ & 0.97 & 0.89 \\
    \hline
  VL$_{220}$ & 3GB $(q_1 = 42.5, q_2=38 {\rm MeV})$ & 36 GeV & 8.9 & $2.00 \times 10^{-39} {\rm cm}^2$ & 0.90 & 0.90 \\
    \hline
  VL$_{270}$ & 3GB $(q_1 = 50, q_2=37.5  {\rm MeV})$ & 32 GeV & 10.3 & $2.10 \times 10^{-39} {\rm cm}^2$ & 0.94 & 0.95  \\
    \hline \hline
   ''  & '' & '' & 14.9  & $1.79 \times 10^{-39} {\rm cm}^2$ & 0.90 & 0.90 \\ 
    \hline
  \end{tabular}
\end{center}
  \caption{ Our three benchmark points. We also indicate what the constraints
on the 3GB(VL$_{270}$) benchmark point would be if a slightly smaller overall
cross-section $\sigma_{p,100{\rm MeV}}$ were chosen.}
\label{tab:bench}
\end{table}

\begin{figure}[t!]
\begin{center}
\includegraphics[width=0.32\textwidth]{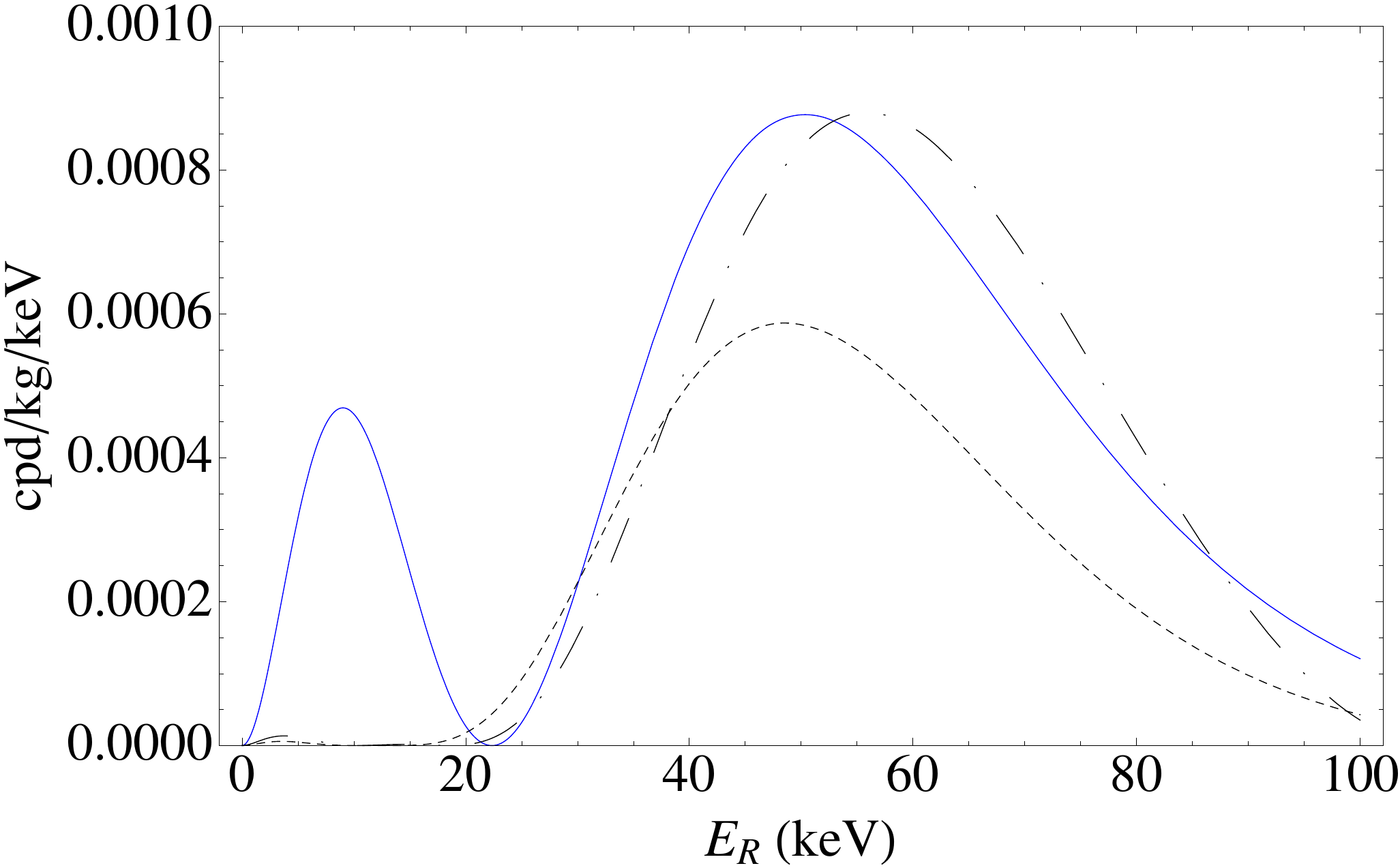}
\includegraphics[width=0.32\textwidth]{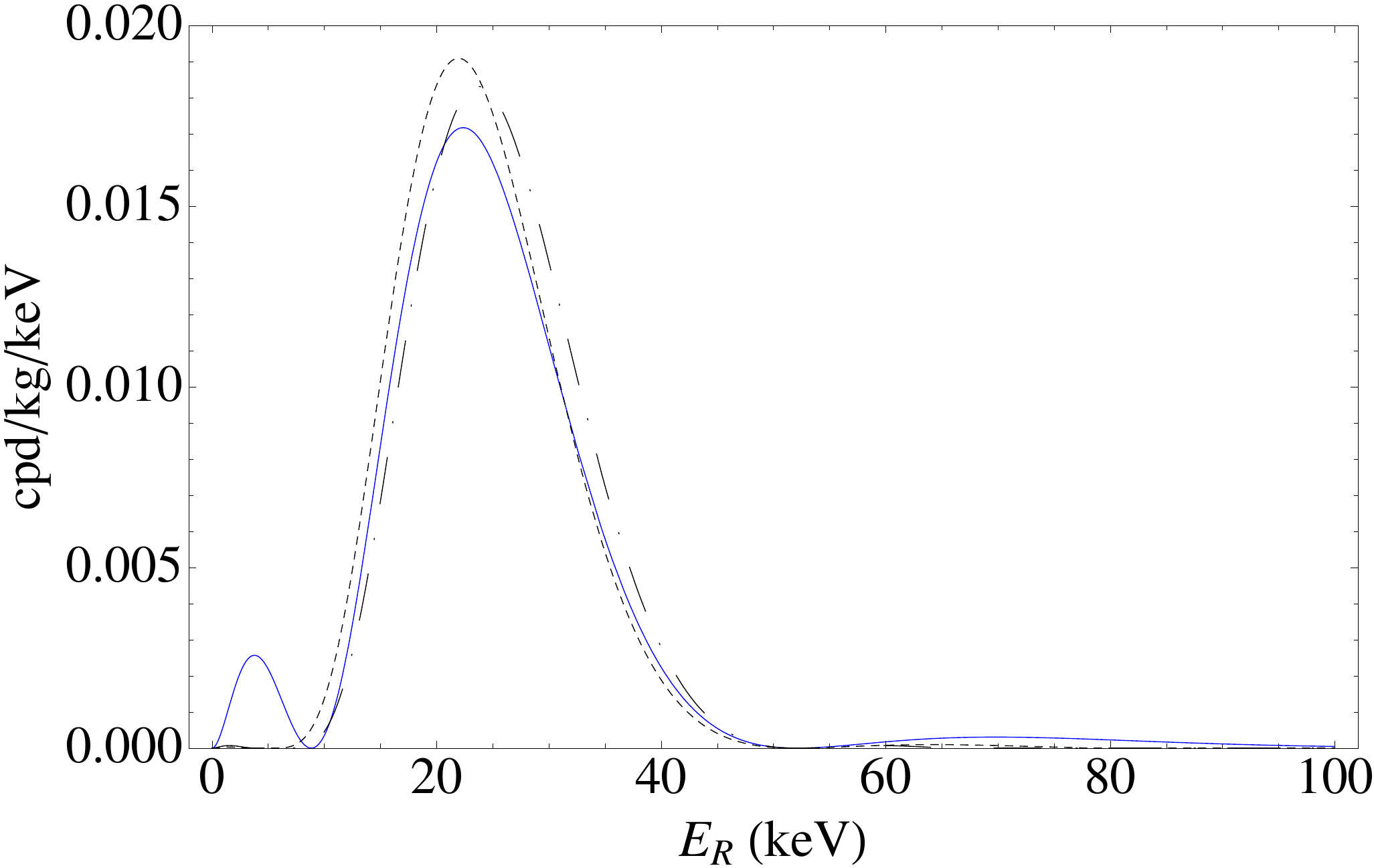}
\includegraphics[width=0.32\textwidth]{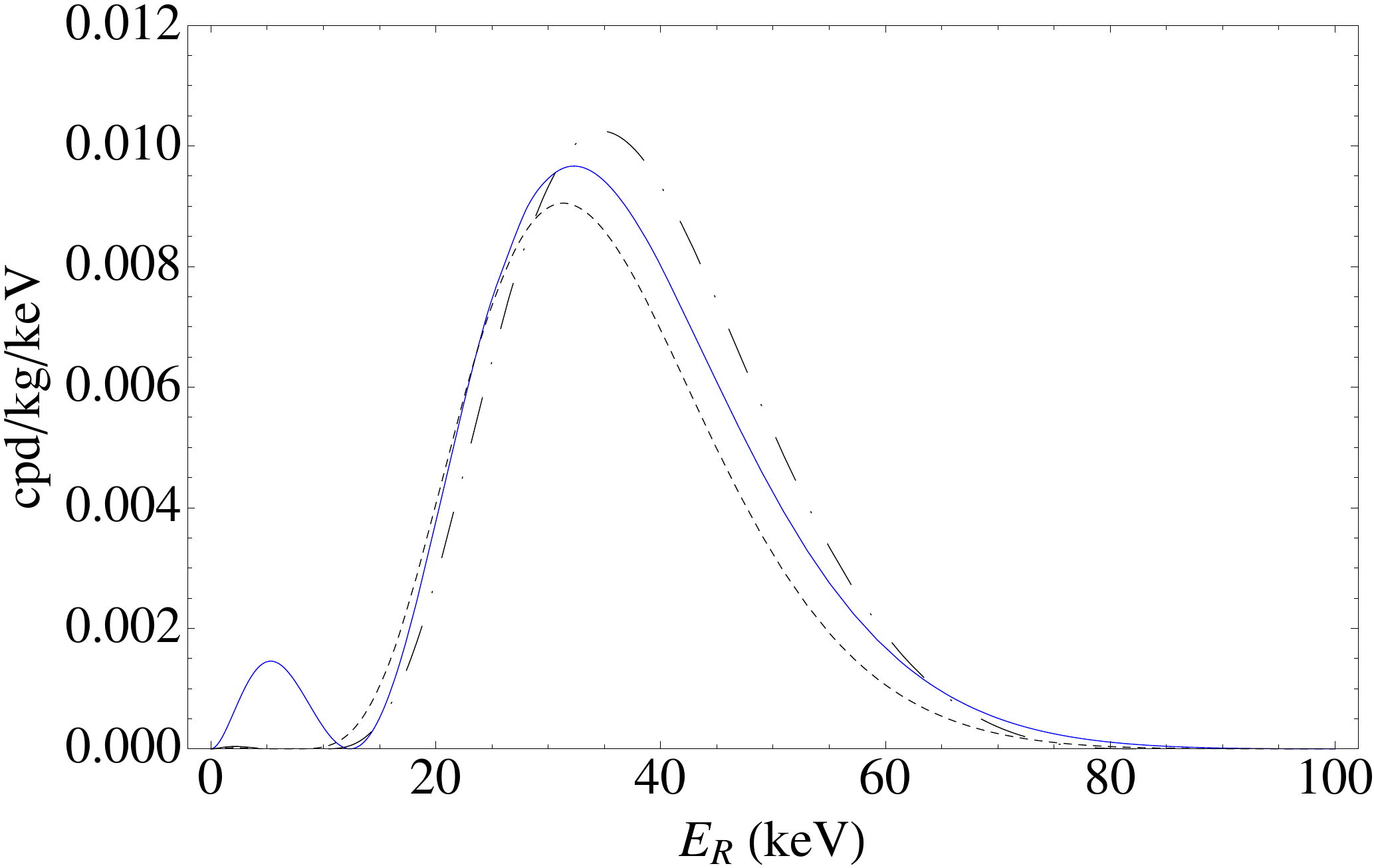}
\caption{ Predicted spectra at Ge (left), W (center) and
Xe (right) for the three
benchmark points in table \ref{tab:bench}.  We have chosen the time of year
to correspond to the dates of CDMS, CRESST-II '07, and ZEPLIN-III, 
respectively.  The 2GB(VL$_{220}$), 
3GB(VL$_{220}$), and 
3GB(VL$_{270}$) models are shown in solid, dashed, and dot-dashed.  }
\label{fig:nullspectra}
\end{center}
\end{figure}

The value of $\chi^2$ for the DAMA 90\% and 99\% contours depends
on the minimum $\chi^2$ for each plot, so it can be helpful to
know this value.  For the 2GB (3GB) model, $\chi^2_{\rm min}
=
4.6, 4.7, 4.7 (9.8, 5.0,5.24)$ for the standard halo, VL$_{220}$, and VL$_{270}$
plot, respectively.  In all cases, $\chi^2_{\rm min}$ is below
$10=12-2$, as appropriate for 12 DAMA bins and 2 free parameters
($\sigma_p$ and $q_0$).  Given the strong dependence of the fits on
$m_{\rm DM}$, it is possible that the $\chi^2$ distribution for our
models more closely follows a distribution with three free parameters,
in which case our $\chi^2_{\rm min}$ is still sufficiently small
at VL$_{220}$, VL$_{270}$, but the DAMA contours will be larger.  In this sense also,
our analysis is conservative.

In table \ref{tab:bench}, we show three benchmark models, one for
the 2GB model with the VL$_{220}$ and two for the 3GB, one for
each VL$_x$ model, along with their statistical constraints
from CDMS and CRESST-II. Note that the null constraints 
are within 97\% for the 2GB model and within 95\%(90\%) for the 
3GB, VL$_{270}$(VL$_{220}$) model.
We have also shown in table \ref{tab:bench} the constraints
on the 3GB(VL$_{270})$ model using a slightly larger value for the
cross-section $\sigma_{p,100{\rm MeV}}$, because such a cross-section
would be permitted by a goodness-of-fit $\chi^2$ test.  More precisely,
 $\chi^2$ for the fit to DAMA in the new 3GB(VL$_{270})$ model is
14.9, which corresponds to a probability of 86.4\% for 10 degrees of
freedom, and 90.6\% for 9 degrees of freedom.
It is, however, just outside the 99\% contour $\chi^2=\chi^2_{\rm min}
+9.21 = 14.4$, which underscores that the constraints are sensitive
to the statistical assumptions being made. 
Finally, we also show in Fig. \ref{fig:nullspectra}
the predicted spectra for our three benchmark points at CDMS and CRESST-II.

\section{Discussion and Future Directions}
\label{sec:conclusions}

In this paper, we have discussed the possibility of using a dynamical form
factor to reconcile the DAMA annual modulation signature with the null results
of other direct detection experiments. \ We have shown that, independent of
specific model assumptions, such a form factor can in fact be consistent with
all available data. \ The picture works to varying degrees depending on the
assumptions one makes concerning the distribution of velocities in the dark
matter Halo.

An important question is how one might distinguish this scenario
from that of inelastic dark matter. \ For this purpose, probably the most
important indicator would be in the energy spectra to be seen at future runs
of the various direct detection experiments.

As discussed in section \ref{sec:qoverlap}, independent of model details, form
factor dark matter leads to a prediction for the events to be seen at any
direct detection experiment within the range of momenta probed by DAMA. \
Although this prediction depends somewhat on the dark matter mass, some
non-trivial event rates will always be present. \ Inelastic dark matter, on
the other hand, has a bit more flexibility in this regard: \ due to the form 
eq. (\ref{eq:vmin})
of the minimum velocity for inelastic scattering, it is possible to choose
parameters such that $v_{\rm min}$ is always above the escape velocity 
at CDMS but below the escape velocity at DAMA in the energy region
of the modulation signal \cite{Neal1,SpencerNeal}.  
The result is that inelastic dark matter may
be consistent with possibilities such as zero events at future runs of CDMS, a
situation which would rule out a pure form factor.

In contrast to this, there are certain aspects of the form factor
setup which are actually more flexible than in the inelastic case: \ inelastic
dark matter makes a firm prediction that all experiments should find zero
events at momenta smaller than a particular cutoff $q_{\min} \sim
\frac{\delta}{v_{\tmop{esc}} + v_e}$, which must be
set approximately to the lower end of the DAMA spectrum. \ A general form 
factor, on the other
hand, could accomodate such events if they were to be seen.

Further differences between the predictions of inelastic and form
factor dark matter could become relevant after large amounts of data have been
collected. \ In particular, details of the energy dependence of the fraction
of modulating events could be used to differentiate the two pictures, but only
with large amounts of statistics. \ It is also possible that future
directional detection experiments could provide an additional handle \cite{Crucis}. \
Finally, although collider signatures for the general form factor scenario are
model dependent, the specific setups discussed in section \ref{sec:models} could lead to
events at low energy $e^+ e^-$ colliders of the types discussed in
\cite{Natalia1, Natalia2}.

There are a wide variety of future directions to pursue within the
general framework we have discussed in this paper. \ Of course, we hope that
future simulations of the galactic halo, including baryons, will yield a more
precise picture for the structure of the dark matter velocity distribution. \
At the present level of understanding, uncertainties in this distribution have
a significant impact on the possible viability and allowed parameter ranges
for not only form factor dark matter, but for essentially any proposed
explanation for the DAMA annual modulation signature.

From the particle physics side, it would certainly be interesting
to try to write down additional types of models leading to form factors with
the general features discussed in section \ref{sec:qoverlap}. \ In this paper we considered a
simple ``proof of principle'', but other model building directions should be
viable as well. \ Within the context of the specific constructions
presented here, it would be worthwhile to try to flesh out a more detailed
structure for the composite dark matter particles, along with a corresponding
cosmological history.

Finally, considering form factor dark matter in the context of a
possible ``channeling'' effect \cite{Channel1,Channel2} 
in the DAMA/NaI detector seems to be a very
promising direction, and will be the subject of a future paper \cite{treepee}.

\

Note added: After this paper appeared on arXiv, there appeared
\cite{Chang:2009yt}, which considers certain dark matter form factors
in the context of spin-dependent cross-sections.

\section*{Acknowledgments}
We have benefited from conversations with A. Cohen, D. Finkbeiner, 
M. McCullough, B. Tweedie, J. Wacker, and N. Weiner.  We especially would like to thank 
M. Kuhlen for comments on uncertainties regarding modeling the dark matter
halo.  
BF is supported by DOE grant DE-FG02-01ER-40676, ALF
is supported by DOE grant DE-FG02-01ER-40676 and NSF CAREER grant
PHY-0645456, and EK is supported by DOE grant DE-FG02-01ER-40676,
NSF CAREER grant PHY-0645456, and an Alfred P. Sloan Fellowship.


\parskip 0pt

\begin{thebibliography}{nn}

\bibitem{DAMALIBRA}
  R.~Bernabei {\it et al.}  [DAMA Collaboration],
  ``First results from DAMA/LIBRA and the combined results with DAMA/NaI,''
  Eur.\ Phys.\ J.\  C {\bf 56}, 333 (2008)
  [arXiv:0804.2741 [astro-ph]].
  %%CITATION = EPHJA,C56,333;%%


\bibitem{DAMA}
  R.~Bernabei {\it et al.}  [DAMA Collaboration],
  ``Search for WIMP annual modulation signature: Results from DAMA / NaI-3 and
  DAMA / NaI-4 and the global combined analysis,''
  Phys.\ Lett.\  B {\bf 480}, 23 (2000).
  %%CITATION = PHLTA,B480,23;%%


\bibitem{fairbairnschwetz}
  M.~Fairbairn and T.~Schwetz,
  ``Spin-independent elastic WIMP scattering and the DAMA annual modulation
  signal,''
  JCAP {\bf 0901}, 037 (2009)
  [arXiv:0808.0704 [hep-ph]].
  %%CITATION = JCAPA,0901,037;%%

\bibitem{Neal1}
  D.~Tucker-Smith, N.~Weiner,
  ``Inelastic Dark Matter,''
  Phys.\ Rev.\  D {\bf 64}, 043502 (2001)
  [arXiv:hep-ph/].
  %%CITATION = PHRVA,D64,043502;%%

\bibitem{SpencerNeal}
  S.~Chang, G.~D.~Kribs, D.~Tucker-Smith and N.~Weiner,
  %``Inelastic Dark Matter in Light of DAMA/LIBRA,''
  Phys.\ Rev.\  D {\bf 79}, 043513 (2009)
  [arXiv:0807.2250 [hep-ph]].
  %%CITATION = PHRVA,D79,043513;%%

\bibitem{JMR}
  J.~March-Russell, C.~McCabe and M.~McCullough,
  ``Inelastic Dark Matter, Non-Standard Halos and the DAMA/LIBRA Results,''
  JHEP {\bf 0905}, 071 (2009)
  [arXiv:0812.1931 [astro-ph]].
  %%CITATION = JHEPA,0905,071;%%

\bibitem{lisapoland}
  Y.~Cui, D.~E.~Morrissey, D.~Poland and L.~Randall,
  ``Candidates for Inelastic Dark Matter,''
  JHEP {\bf 0905}, 076 (2009)
  [arXiv:0901.0557 [hep-ph]].
  %%CITATION = JHEPA,0905,076;%%

\bibitem{Crucis}
  D.~P.~Finkbeiner, T.~Lin and N.~Weiner,
  ``Inelastic Dark Matter and DAMA/LIBRA: An Experimentum Crucis,''
  arXiv:0906.0002 [astro-ph.CO].
  %%CITATION = ARXIV:0906.0002;%%

\bibitem{Winkler}
  K.~Schmidt-Hoberg and M.~W.~Winkler,
  ``Improved Constraints on Inelastic Dark Matter,''
  arXiv:0907.3940 [astro-ph.CO].
  %%CITATION = ARXIV:0907.3940;%%

%\cite{Lewin:1995rx}
\bibitem{lewinsmith}
  J.~D.~Lewin and P.~F.~Smith,
  ``Review of mathematics, numerical factors, and corrections for dark  matter
  experiments based on elastic nuclear recoil,''
  Astropart.\ Phys.\  {\bf 6}, 87 (1996).
  %%CITATION = APHYE,6,87;%%


\bibitem{lewinsmithlong}
  P.~F.~Smith and J.~D.~Lewin,
  ``Dark Matter Detection,''
  Phys.\ Rept.\  {\bf 187}, 203 (1990).
  %%CITATION = PRPLC,187,203;%%

%\cite{Duda:2006uk}
\bibitem{Duda:2006uk}
  G.~Duda, A.~Kemper and P.~Gondolo,
  ``Model independent form factors for spin independent neutralino nucleon
  scattering from elastic electron scattering data,''
  JCAP {\bf 0704}, 012 (2007)
  [arXiv:hep-ph/0608035].
  %%CITATION = JCAPA,0704,012;%%


\bibitem{fricke}
  G.~Fricke et al.,
  ``Nuclear Ground State Charge Radii from Electromagnetic Interactions,''
   Atomic Data and Nuclear Data Tables {\bf 60}, 177 (1995).

\bibitem{vesc}
  M.~C.~Smith {\it et al.},
  %``The RAVE Survey: Constraining the Local Galactic Escape Speed,''
  Mon.\ Not.\ Roy.\ Astron.\ Soc.\  {\bf 379}, 755 (2007)
  [arXiv:astro-ph/0611671].
  %%CITATION = MNRAA,379,755;%%

\bibitem{Feng:2008dz}
  J.~L.~Feng, J.~Kumar and L.~E.~Strigari,
  ``Explaining the DAMA Signal with WIMPless Dark Matter,''
  Phys.\ Lett.\  B {\bf 670}, 37 (2008)
  [arXiv:0806.3746 [hep-ph]].
  %%CITATION = PHLTA,B670,37;%%

\bibitem{Petriello:2008jj}
  F.~Petriello and K.~M.~Zurek,
  ``DAMA and WIMP dark matter,''
  JHEP {\bf 0809}, 047 (2008)
  [arXiv:0806.3989 [hep-ph]].
  %%CITATION = JHEPA,0809,047;%%

\bibitem{Bottino:2008mf}
  A.~Bottino, F.~Donato, N.~Fornengo and S.~Scopel,
  ``Interpreting the recent results on direct search for dark matter particles
  in terms of relic neutralino,''
  Phys.\ Rev.\  D {\bf 78}, 083520 (2008)
  [arXiv:0806.4099 [hep-ph]].
  %%CITATION = PHRVA,D78,083520;%%

\bibitem{Foot:2008nw}
  R.~Foot,
  ``Mirror dark matter and the new DAMA/LIBRA results: A simple explanation for
  a beautiful experiment,''
  Phys.\ Rev.\  D {\bf 78}, 043529 (2008)
  [arXiv:0804.4518 [hep-ph]].
  %%CITATION = PHRVA,D78,043529;%%

\bibitem{Gondolo:2005hh}
  P.~Gondolo and G.~Gelmini,
  ``Compatibility of DAMA dark matter detection with other searches,''
  Phys.\ Rev.\  D {\bf 71}, 123520 (2005)
  [arXiv:hep-ph/0504010].
  %%CITATION = PHRVA,D71,123520;%%

\bibitem{SpencerAaronNeal}
  S.~Chang, A.~Pierce and N.~Weiner,
  ``Using the Energy Spectrum at DAMA/LIBRA to Probe Light Dark Matter,''
  arXiv:0808.0196 [hep-ph].
  %%CITATION = ARXIV:0808.0196;%%

\bibitem{Yellin}
  S.~Yellin,
  %``Finding an upper limit in the presence of unknown background,''
  Phys.\ Rev.\  D {\bf 66}, 032005 (2002)
  [arXiv:physics/0203002].
  %%CITATION = PHRVA,D66,032005;%%

\bibitem{vialacteaI}
  J.~Diemand, M.~Kuhlen and P.~Madau,
  ``Dark matter substructure and gamma-ray annihilation in the Milky Way
  halo,''
  Astrophys.\ J.\  {\bf 657}, 262 (2007)
  [arXiv:astro-ph/0611370].
  %%CITATION = ASJOA,657,262;%%


\bibitem{CDMS}
  D.~S.~Akerib {\it et al.}  [CDMS Collaboration],
  ``First results from the cryogenic dark matter search in the Soudan
  Underground Lab,''
  Phys.\ Rev.\ Lett.\  {\bf 93}, 211301 (2004)
  [arXiv:astro-ph/0405033].
  %%CITATION = PRLTA,93,211301;%%

\bibitem{CDMS2}
  D.~S.~Akerib {\it et al.}  [CDMS Collaboration],
  ``Limits on spin-independent WIMP nucleon interactions from the two-tower
  run of the Cryogenic Dark Matter Search,''
  Phys.\ Rev.\ Lett.\  {\bf 96}, 011302 (2006)
  [arXiv:astro-ph/0509259].
  %%CITATION = PRLTA,96,011302;%%

\bibitem{CDMSFIVE}
  Z.~Ahmed {\it et al.}  [CDMS Collaboration],
  ``Search for Weakly Interacting Massive Particles with the First Five-Tower
  Data from the Cryogenic Dark Matter Search at the Soudan Underground
  Laboratory,''
  Phys.\ Rev.\ Lett.\  {\bf 102}, 011301 (2009)
  [arXiv:0802.3530 [astro-ph]].
  %%CITATION = PRLTA,102,011301;%%

\bibitem{CRESST2004}
  G.~Angloher {\it et al.},
  ``Limits on WIMP dark matter using scintillating CaWO-4 cryogenic  detectors
  with active background suppression,''
  Astropart.\ Phys.\  {\bf 23}, 325 (2005)
  [arXiv:astro-ph/0408006].


\bibitem{CRESST2}
  G.~Angloher {\it et al.},
  ``Commissioning Run of the CRESST-II Dark Matter Search,''
  arXiv:0809.1829 [astro-ph].
  %%CITATION = ARXIV:0809.1829;%%

\bibitem{Lang:2009ge}
  R.~F.~Lang and W.~Seidel,
  ``Search for Dark Matter with CRESST,''
  arXiv:0906.3290 [astro-ph.IM].

\bibitem{XENON10}
  J.~Angle {\it et al.}  [XENON Collaboration],
  ``First Results from the XENON10 Dark Matter Experiment at the Gran Sasso
  National Laboratory,''
  Phys.\ Rev.\ Lett.\  {\bf 100}, 021303 (2008)
  [arXiv:0706.0039 [astro-ph]].
  %%CITATION = PRLTA,100,021303;%%

\bibitem{XENON10IDM}
  X.~Collaboration,
  ``Constraints on inelastic dark matter from XENON10,''
  arXiv:0910.3698 [astro-ph.CO].
  %%CITATION = ARXIV:0910.3698;%%

\bibitem{ZEPLIN2}
  G.~J.~Alner {\it et al.},
  ``First limits on WIMP nuclear recoil signals in ZEPLIN-II: A two phase xenon
  detector for dark matter detection,''
  Astropart.\ Phys.\  {\bf 28}, 287 (2007)
  [arXiv:astro-ph/0701858].
  %%CITATION = APHYE,28,287;%%



\bibitem{ZEPLIN3}
  V.~N.~Lebedenko {\it et al.},
  ``Result from the First Science Run of the ZEPLIN-III Dark Matter Search
  Experiment,''
  arXiv:0812.1150 [astro-ph].
  %%CITATION = ARXIV:0812.1150;%%

\bibitem{KIMS}
  S.~K.~Kim  [KIMS Collaboration],
  ``New results from the KIMS experiment,''
  J.\ Phys.\ Conf.\ Ser.\  {\bf 120}, 042021 (2008).
  %%CITATION = 00462,120,042021;%%


\bibitem{Natalia1}
  J.~D.~Bjorken, R.~Esseg, P.~Schuster, N.~Toro,
  ``New Fixed-Target Experiments to Search for Dark Gauge Forces,''
  [arXiv:0906.0580 [hep-ph]].

\bibitem{Natalia2}
  R.~Esseg, P.~Schuster, N.~Toro,
  ``Probing Dark Forces and New Hidden Sectors at Low-Energy e+e- Colliders,''
  [arXiv:0903.3941 [hep-ph]].



\bibitem{Channel1}
  E.~M.~Drobyshevski,
  ``Channeling Effect and Improvement of the Efficiency of Charged Particle
  %Registration with Crystal Scintillators,''
  Mod.\ Phys.\ Lett.\  A {\bf 23}, 3077 (2008)
  [arXiv:0706.3095 [physics.ins-det]].
  %%CITATION = MPLAE,A23,3077;%%

\bibitem{Channel2}
  R.~Bernabei {\it et al.},
  ``Possible implications of the channeling effect in NaI(Tl) crystals,''
  Eur.\ Phys.\ J.\  C {\bf 53}, 205 (2008)
  [arXiv:0710.0288 [astro-ph]].
  %%CITATION = EPHJA,C53,205;%%


\bibitem{treepee}
  B.~Feldstein, A.~L.~Fitzpatrick, E.~Katz and B.~Tweedie,
  ``A Simple Explanation for DAMA with Moderate Channeling,''
  arXiv:0910.0007 [hep-ph]
%\bibitem{treepee}
%  B.~Feldstein, A.~L.~Fitzpatrick, E.~Katz, B.~Tweedie,
%  in preparation.

\bibitem{Chang:2009yt}
  S.~Chang, A.~Pierce and N.~Weiner,
  ``Momentum Dependent Dark Matter Scattering,''
  arXiv:0908.3192 [hep-ph].
  %%CITATION = ARXIV:0908.3192;%%

\end{thebibliography}
\end{document}